\documentclass{sig-alternate}

\usepackage{mathptmx} 
\usepackage{comment}

\newcommand*\mycirc[1]
{\tikz[baseline=(char.base)]{
  \node[shape=circle,draw,inner sep=0.5pt,fill=black,text=white,font=\sffamily] (char) {#1};}}

\usepackage[export]{adjustbox}

\newcommand{\ignore}[1]{}
\usepackage{fancyhdr}
\usepackage[normalem]{ulem}

\usepackage{xspace}
\usepackage{color}
\usepackage{graphicx}
\usepackage{url}
\usepackage[nocompress]{cite}

\usepackage[font=small]{subfig}

\usepackage{algorithm}
\usepackage{algorithmic}
\usepackage{wrapfig}
\usepackage{tikz}
\usepackage{amsmath}
\usepackage{booktabs}
\usepackage{paralist}
\usepackage[hang,flushmargin]{footmisc} 
\usepackage{cleveref}

\newboolean{publicversion}
\setboolean{publicversion}{true}

\ifthenelse{\boolean{publicversion}}{
  \newcommand{\grumbler}[2]{}
}{
  \newcommand{\grumbler}[2]{\textcolor{red}{\bf #1: #2}}
}

\newcommand{\cjr}[1]{\grumbler{CJR}{#1}}
\newcommand{\vm}[1]{\grumbler{VanceM}{#1}}
\newcommand{\rachata}[1]{\grumbler{Rachata}{#1}}

\newcommand{\para}[1]{\vspace{3pt}\noindent\textbf{#1}}

\crefformat{footnote}{#2\footnotemark[#1]#3}

\newcommand{\name}{\textit} 
\newcommand{\titleShort}[0]{\name{MASK}\xspace}
\newcommand{\tlbtokenname}[0]{\name{TLB-Fill Tokens}\xspace}
\newcommand{\cachebypass}[0]{\name{TLB-Request-Aware L2 Bypass}\xspace}
\newcommand{\dramsched}[0]{\name{Address-Space-Aware DRAM Scheduler}\xspace}

\newcommand{\goldQ}[0]{\name{Golden Queue}\xspace}
\newcommand{\silverQ}[0]{\name{Silver Queue}\xspace}
\newcommand{\normalQ}[0]{\name{Normal Queue}\xspace}

\newcommand{\titleLong}[0]{Multi-Address Space Concurrent Kernels\xspace}
\newcommand{\titleLongEmph}[0]{\textbf{M}ulti-\textbf{A}ddress \textbf{S}pace Concurrent \textbf{K}ernels\xspace}

\date{}

\title{Improving Multi-Application Concurrency Support Within the GPU Memory System}

\author{
Rachata Ausavarungnirun$\star$ \qquad Christopher J. Rossbach$\dagger\ddagger$ \qquad Vance Miller$\dagger$ \\ Joshua Landgraf$\dagger$ \qquad Saugata Ghose$\star$ \qquad Jayneel Gnadhi$\ddagger$ \\ \qquad Adwait Jog$\S$ \qquad Onur Mutlu$\ast$$\star$\\ 
\\
\normalsize
    $\star$Carnegie Mellon University \qquad $\dagger$University of Texas at Austin \qquad $\ddagger$VMware Research Group\\
\normalsize $\S$College of William and Mary \qquad $\ast$ETH Zurich
}

\begin{document}

\setlength{\abovedisplayskip}{0pt}
\setlength{\belowdisplayskip}{1pt}

\maketitle
\pagestyle{plain}

\begin{abstract}

GPUs exploit a high degree of thread-level parallelism to efficiently
hide long-latency stalls. Thanks to their 
latency-hiding abilities and continued improvements in programmability,
GPUs are becoming a more essential computational resource.
Due to the heterogeneous compute requirements of different
applications, there is a growing need to \emph{share} the GPU across multiple
applications in large-scale computing environments.
However, while CPUs offer relatively seamless multi-application concurrency, and are
an excellent fit for multitasking and for virtualized environments, GPUs 
currently offer only primitive support for multi-application concurrency.

Much of the problem in a contemporary GPU lies within the memory system, where
multi-application execution requires virtual memory support to manage the
address spaces of each application and to provide memory protection.
In this work, we perform a detailed analysis of the major problems in state-of-the-art
GPU virtual memory management that hinders multi-application execution.
Existing GPUs are designed to share memory between the CPU and GPU, but do not
handle multi-application support \emph{within the GPU} well.  We find that when
multiple applications spatially share the GPU, there is a significant amount of
inter-core thrashing on the shared TLB within the GPU.
The TLB contention is high enough to prevent the GPU from successfully hiding
stall latencies, thus becoming a first-order performance concern.

Based on our analysis, we introduce \titleShort{}, 
a memory hierarchy design that provides low-overhead virtual memory
support for the concurrent execution of multiple applications.
\titleShort{} extends the GPU 
memory hierarchy to efficiently support address translation
through the use of multi-level TLBs, and uses translation-aware memory and
cache management to maximize throughput in the presence of 
inter-application contention. \titleShort{} uses a novel 
token-based approach to reduce TLB miss overheads, and its
L2 cache bypassing mechanisms and application-aware memory 
scheduling reduce the interference between address translation and 
data requests. 
\titleShort{} restores much of the thread-level parallelism that was previously
lost due to address translation. 
Relative to a state-of-the-art GPU TLB,
\titleShort{} improves system throughput by 45.2\%, improves IPC throughput
by 43.4\%, reduces unfairness by 22.4\%, and \titleShort{} performs within 23\%
of the ideal design with no translation overhead. 

\end{abstract}

\section{Introduction}

Graphics Processing Units (GPUs) provide high throughput by exploiting a high
degree of thread-level parallelism.  A GPU executes a group of threads (i.e., a
\emph{warp}) in lockstep (i.e., each thread in the warp executes the same
instruction concurrently).  When a warp stalls, the GPU hides the latency
of this stall by scheduling and executing another warp.
The use of GPUs to accelerate general-purpose GPU
(GPGPU) applications has become common practice, in large part due to the
large performance improvements that GPUs provide for applications in
diverse domains~\cite{flynn, rodinia, parboil, mars, lonestar}.  The compute
density of GPUs continues to grow, with GPUs expected to provide as many as
128 streaming multiprocessors per chip in the near 
future~\cite{arunkumar-isca17, vijay-hpca17}.  While the increased 
compute density can help many GPGPU applications, it exacerbates the growing
need to \emph{share} the GPU streaming multiprocessors across multiple
applications.  This is especially true in large-scale computing environments,
such as
cloud servers, where a diverse range
of application requirements exists.  In order to enable efficient GPU hardware
utilization in the face of application heterogeneity, these large-scale
environments rely on the ability to virtualize the compute 
resources and execute multiple applications concurrently~\cite{ept, npt, 
amd-io-virt, intel-io-virt}.  

The adoption of discrete GPUs in large-scale computing environments is hindered
by the primitive virtualization support in contemporary GPUs.  While hardware 
virtualization support has improved for integrated GPUs~\cite{kaveri}, the 
current virtualization support for discrete GPUs is insufficient, even though 
discrete GPUs provide the highest available compute density and remain the 
platform of choice in many domains~\cite{tensorflow}. 
Two alternatives for discrete virtualization are time multiplexing and spatial
multiplexing.
Emerging GPU architectures support time multiplexing the GPU by providing
application preemption~\cite{lindholm,pascal}, but this support currently
does not scale well with the number of applications. Each additional
application introduces a high degree of contention for the GPU resources (Section~\ref{s:time-multiplex}).
Spatial multiplexing allows us to share a GPU among applications much as we
currently share CPUs, by providing support for \emph{multi-address-space 
concurrency} (i.e., the concurrent execution of kernels from
different processes or guest VMs).  By efficiently and dynamically managing the
kernels that execute concurrently on the GPU, spatial multiplexing avoids the 
scaling issues of time multiplexing.  
To support spatial multiplexing, GPUs must provide architectural support for
memory virtualization and memory protection domains.

\sloppypar{The architectural support for spatial multiplexing in contemporary GPUs is not
well-suited for concurrent multi-application execution.
Recent efforts at improving address translation support within
GPUs~\cite{powers-hpca14, pichai-asplos14, tianhao-hpca16, abhishek-ispass16,cong-hpca17} 
eschew MMU-based or IOMMU-based~\cite{amd-io-virt,amit10wiosca}
address translation in favor of TLBs close to shader cores. 
These works do not explicitly target concurrent multi-application execution
\emph{within the GPU}, and are instead focused on unifying the CPU and GPU
memory address spaces~\cite{amd-hsa}.
We perform a thorough analysis of concurrent multi-application execution when
these state-of-the-art address translation techniques are employed within a state-of-the-art
GPU (Section~\ref{sec:motiv}).
We make four \emph{key observations} from our analysis.
First, we find that for concurrent multi-application execution, a shared L2 TLB
is more effective than the highly-threaded page table walker and page walk
cache proposed in \cite{powers-hpca14} for the unified CPU-GPU memory address
space.}
Second, for both the shared L2 TLB and the page walk cache, TLB misses become
a major performance bottleneck with concurrent multi-application execution, 
despite the latency-hiding properties of the GPU.
A TLB miss incurs a high latency, as each miss must walk through multiple 
levels of a page table to find the desired address translation.
Third, we observe that a single TLB miss can frequently stall multiple warps at once.
Fourth, we observe that contention between applications induces significant thrashing on the 
shared TLB and significant interference between TLB misses and data requests 
throughout the GPU memory system.
Thus, with only a few simultaneous TLB misses, it becomes 
difficult for the GPU to find a warp that can be scheduled for execution, 
defeating the GPU's basic techniques for hiding the latency of stalls. 

Thus, based on our extensive analysis, we conclude that
\emph{address translation becomes a first-order performance concern} in GPUs
when multiple applications are executed concurrently.
\emph{Our goal} in this work is to develop new techniques that can alleviate 
the severe address translation bottleneck existing in state-of-the-art GPUs.

\sloppypar{
To this end, we propose \titleLongEmph{} (\titleShort{}), a new cooperative resource
management framework and TLB design for GPUs that minimizes inter-application interference 
and translation overheads.
\titleShort{} takes advantage of 
locality across shader cores to reduce TLB misses, and relies on
three novel techniques to minimize translation overheads. 
The overarching key idea is to make the entire memory
hierarchy \emph{TLB request aware}.
First, \tlbtokenname provide a
TLB-selective-fill mechanism to reduce thrashing in the shared L2 TLB,
including a bypass cache to increase the TLB hit rate.
Second, a low-cost scheme for selectively bypassing TLB-related requests at
the L2 cache reduces interference between TLB-miss and data requests.
Third, \titleShort{}'s memory scheduler prioritizes TLB-related requests 
to accelerate page table walks. 

The techniques employed by \titleShort{} are highly effective at alleviating 
the address translation bottleneck.  Through the use of TLB-request-aware
policies throughout the memory hierarchy, \titleShort{} ensures that the first
two levels of the page table walk during a TLB miss are serviced quickly.  
This reduces the overall latency of a TLB miss significantly.  Combined with
a significant reduction in TLB misses, \titleShort{} allows the GPU to
successfully hide the latency of the TLB miss through thread-level parallelism.
As a result, \titleShort{} improves system throughput
by 45.2\%, improves IPC throughput by 43.4\%, and reduces unfairness by
22.4\% over a state-of-the-art GPU memory management unit (MMU) design~\cite{powers-hpca14}.
\titleShort{} provides performance within only 23\% of a perfect TLB that
always hits.

This paper makes the following contributions:
\begin{compactitem} 

\item To our knowledge, this is the first work to provide a thorough
analysis of GPU memory virtualization under multi-address-space concurrency, and
to demonstrate the large impact address translation has on latency hiding
within a GPU. We demonstrate a need for new techniques to alleviate
interference induced by multi-application execution. 

\item We design an MMU that is optimized for 
GPUs that are dynamically partitioned spatially across protection domains, rather
than GPUs that are time-shared.

\item We propose \titleShort{}, which consists of three novel techniques that 
increase TLB request awareness across the entire memory hierarchy.
These techniques work together to significantly improve system performance,
IPC throughput, and fairness over a state-of-the-art GPU MMU.


\end{compactitem}

\section{Background}
\label{sec:background}


There has been an emerging need to \emph{share} the GPU
hardware among multiple applications.  As a result, recent work has
enabled support for GPU virtualization, where a single physical GPU can be
shared transparently across multiple applications, with each application having
its own address space.\footnote{In this paper, we use the term \emph{address space} 
to refer to distinct \emph{memory protection domains}, whose access to 
resources must be isolated and protected during GPU virtualization.}  
Much of this work has relied on traditional time and spatial multiplexing techniques that
have been employed by CPUs, and state-of-the-art GPUs currently contain elements
of both types of techniques~\cite{gpuvm,gVirt,vmCUDA}.
Unfortunately, as we discuss in this section, existing GPU virtualization
implementations are too coarse, bake fixed policy into hardware, or leave 
system software without the fine-grained resource management primitives needed to implement 
truly transparent device virtualization.

\subsection{Time Multiplexing}
\label{s:time-multiplex}

Most modern systems time-share GPUs~\cite{kepler,lindholm}. 
These designs are optimized for the case
where \emph{no concurrency exists} between kernels from
different address spaces. This simplifies memory protection and scheduling at 
the cost of two fundamental tradeoffs. First, it results
in underutilization
when kernels from a single address space are unable to
fully utilize all of the GPU's resources~\cite{nmnl-pact13,cpugpu-micro,asplos-sree,wang-hpca16,mafia}.
Second, it limits the ability of a scheduler to provide
forward-progress or QoS guarantees, leaving applications vulnerable to unfairness
and starvation~\cite{ptask}.

While preemption support could 
allow a time-sharing scheduler to avoid pathological unfairness
(e.g., by context switching at a fine granularity), GPU preemption support 
remains an active research area~\cite{isca-2014-preemptive, gebhart}. Software
approaches~\cite{wang-hpca16} sacrifice memory protection. 
NVIDIA's Kepler~\cite{kepler} and Pascal~\cite{pascal} architectures 
support preemption at thread block and instruction granularity respectively.
We find empirically, that neither is well optimized for inter-application
interference.

Figure~\ref{fig:context-switch-nvidia} shows the overhead (i.e., performance loss) \emph{per process} 
when the NVIDIA K40 and GTX 1080 GPUs are contended by multiple application processes.  Each process runs a kernel that interleaves 
basic arithmetic operations with loads and stores into shared and global memory,
with interference from a GPU matrix-multiply program.
The overheads range from 8\% per process for the K40, to 10\% or 12\% for the GTX 1080.
While the performance cost is significant, we also found 
inter-application interference pathologies to be easy to create: for example, a kernel from one process 
consuming the majority of shared memory can easily cause kernels from other processes to fail at dispatch.
While we expect preemption support to improve in future hardware,
we seek a solution that does not depend on it. 

\begin{figure}[h]
\centering
  \includegraphics[width=\columnwidth]{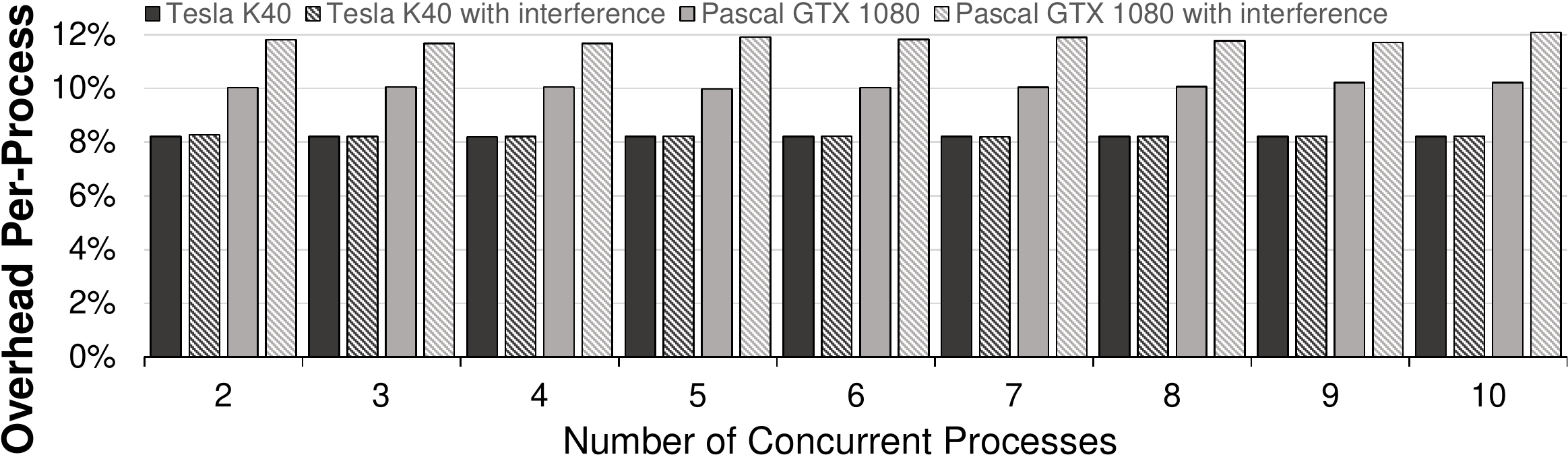}
  \caption{Context switch overheads under contention on K40 and GTX 1080.}
  \label{fig:context-switch-nvidia}
\end{figure}


\subsection{Spatial Multiplexing}

Resource utilization can be improved with {\em spatial
multiplexing}~\cite{gpu-multitasking}, as the ability to execute multiple kernels
\emph{concurrently} enables the system to co-schedule kernels that have complementary
resource demands, and can enable independent progress guarantees for different kernels.
NVIDIA's stream~\cite{kepler}
support, which co-schedules kernels from independent ``streams'' in a single
address space, relies on similar basic concepts, as does application-specific software 
scheduling of multiple kernels in hardware~\cite{asplos-sree, mafia} and
GPU simulators~\cite{nmnl-pact13,sms,cpugpu-micro}.  Software approaches (e.g., Elastic
Kernels~\cite{asplos-sree}) require programmers to manually time-slice kernels
to enable mapping them onto CUDA streams for concurrency. While sharing techniques that leverage
the stream abstraction support flexible demand-partitioning of resources, 
they all share critical drawbacks. When kernels from different applications have
complementary resource demands, the GPU remains underutilized. More importantly,
merging kernels into a single address space sacrifices memory protection, a key 
requirement in virtualized settings. 


Multi-address-space concurrency support can address these
shortcomings by enabling a scheduler to look beyond kernels from the
current address space when resources are under-utilized. Moreover,
QoS and forward-progress guarantees can be enabled by giving
partitions of the hardware simultaneously to kernels from different
address spaces. For example, long-running kernels from one application
need not complete before kernels from another may be dispatched.
NVIDIA and AMD both offer products~\cite{grid,firepro} with
hardware virtualization support for statically partitioning GPUs
across VMs, but even this approach has critical shortcomings. \rachata{Add context switching overhead?} The system 
must select from a handful of different partitioning schemes, determined
at startup, which is fundamentally inflexible. The system cannot adapt 
to changes in demand or mitigate interference, which are key goals 
of virtualization layers.

\section{Baseline Design}
\label{sec:background-spatial}

Our goal in this work is to develop efficient address translation techniques
for GPUs that allow for flexible, fine-grained spatial multiplexing of the GPU
across multiple address spaces, ensuring protection across memory
protection domains.  Our primary focus is 
on optimizing a memory hierarchy design extended with TLBs,
which are used in a state-of-the-art GPU~\cite{powers-hpca14}. 
Kernels running concurrently on different compute units share 
components of the memory hierarchy such as lower level caches, so 
ameliorating contention for those components is an important concern. 
In this section, we explore the performance costs and bottlenecks induced by different 
components in a baseline design for GPU address translation, motivating the need
for \titleShort{}.

\subsection{Memory Protection Support}
\label{sec:memory-protection}

To ensure that memory accesses from kernels running in different address spaces
remain isolated, we make use of TLBs and mechanisms for reducing TLB miss costs within the GPU.
We adopt the current state-of-the-art for GPU TLB design for CPU-GPU heterogeneous systems
proposed by Power et al.~\cite{powers-hpca14}, and extend the design to handle multi-address-space concurrency, 
as shown in Figure~\ref{fig:tlb-powers-hpca14}. Each core has a private L1 cache (\mycirc{1}), and all
cores share a highly-threaded page table walker (\mycirc{2}).  On a TLB miss,
the shared page table walker first probes a page walk cache
(\mycirc{3}).\footnote{In our evaluation, we provision the page walk cache to be
16-way, with 1024 entries.} A miss in the page walk cache goes to the shared L2 cache and (if need be) main memory.

\begin{figure}[h]%
\centering
\resizebox{\columnwidth}{!}{%
\subfloat[TLB design from~\cite{powers-hpca14}]{{
	\includegraphics[height=80pt]{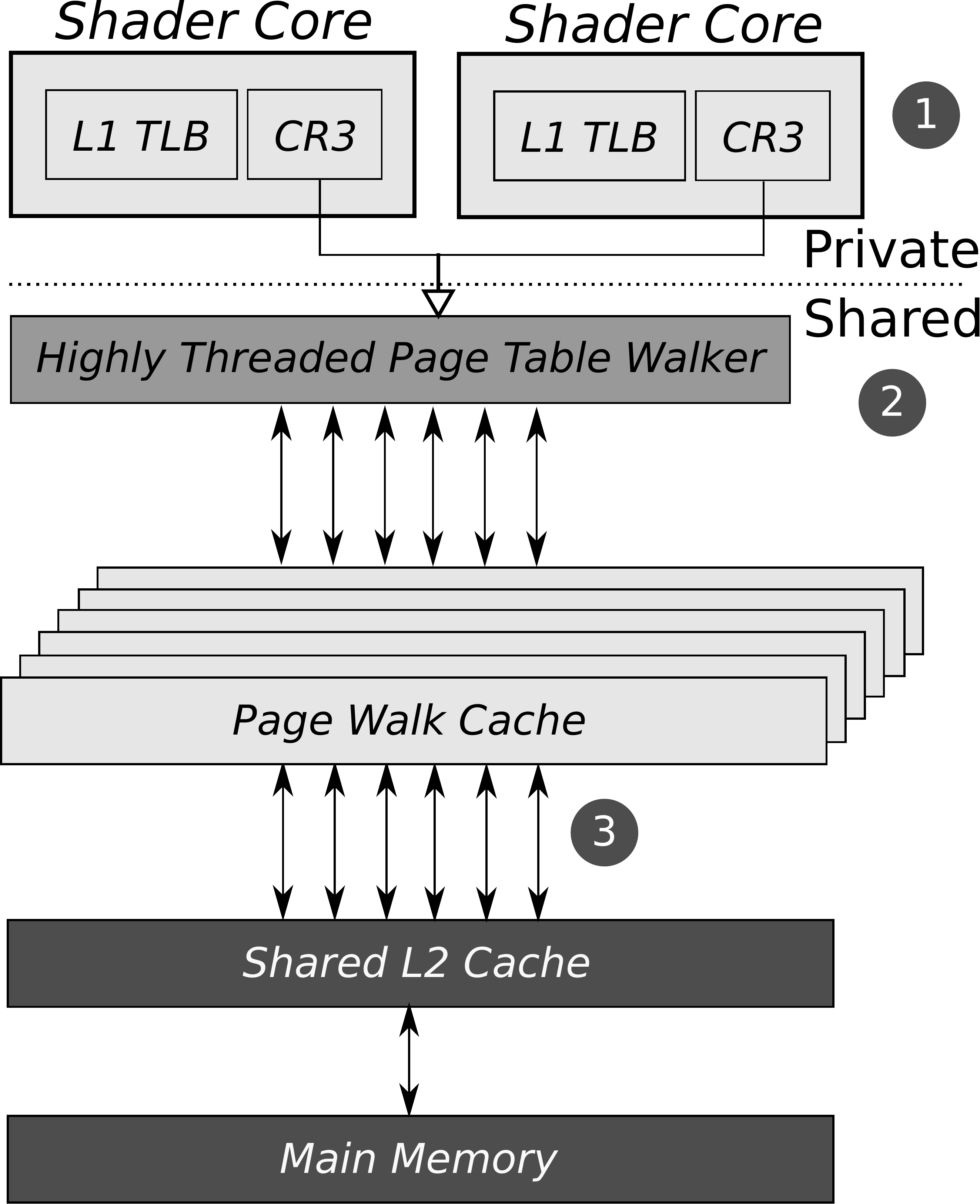}%
	\label{fig:tlb-powers-hpca14}
}}%
\qquad
\subfloat[\titleShort{}'s baseline TLB design.]{{
	\includegraphics[height=80pt]{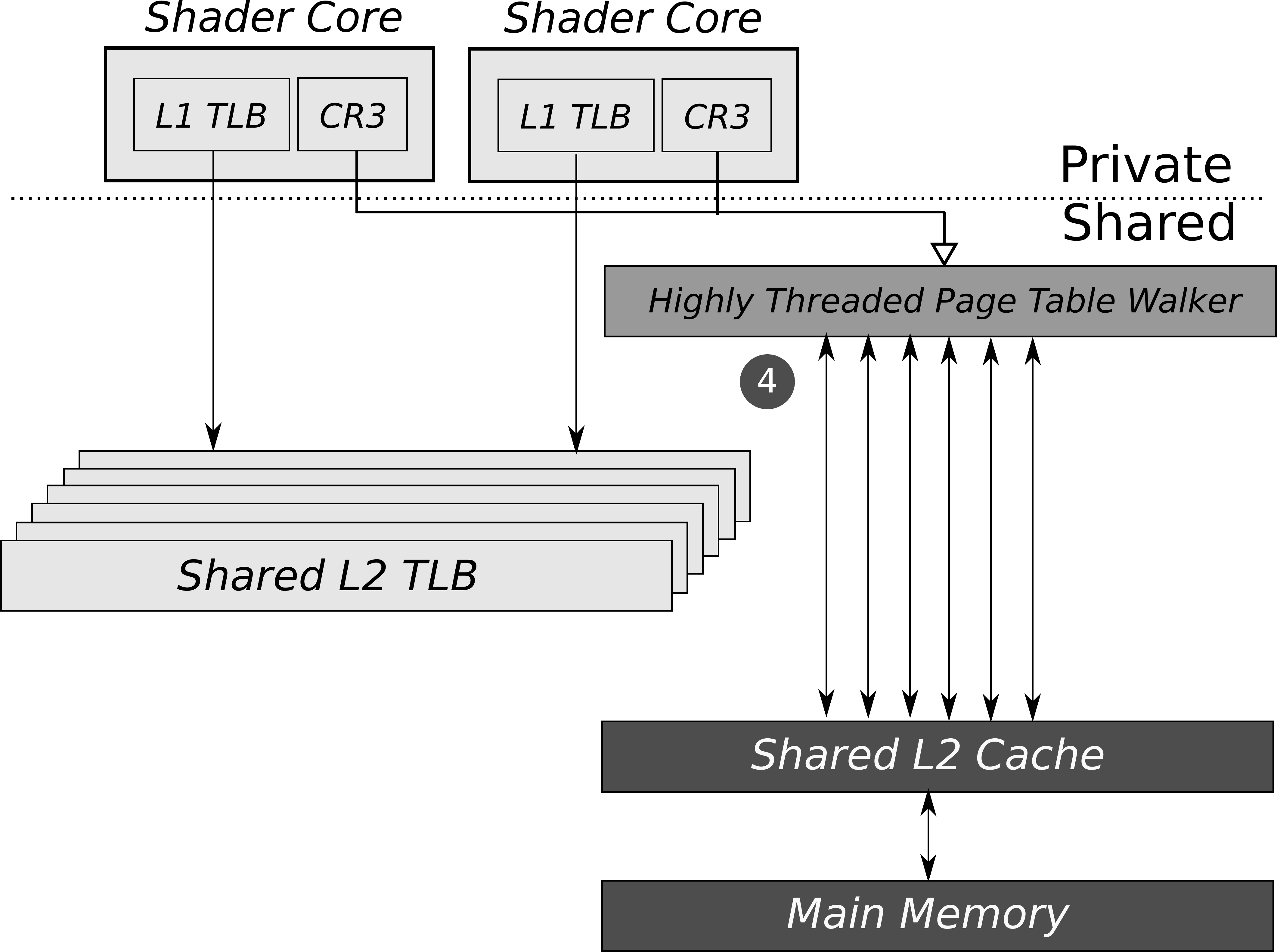}%
	\label{fig:tlb-mask-baseline}
}}%
}
\caption{Baseline TLB designs \vm{Better caption?}}%
\end{figure}

\subsection{Page Walk Caches}
\label{sec:pwc}

Techniques to avoid misses and hide or reduce their latency 
are well-studied in the literature. 
To conserve space, we do not discuss the combinations of techniques that we considered,
and focus on the 
design which we ultimately selected as the baseline for \titleShort{}, shown in Figure~\ref{fig:tlb-mask-baseline}. 
The design differs
from~\cite{powers-hpca14} by eliminating the page walk cache, and
instead dedicating the same chip area to 1)~a shared L2 TLB with
entries extended with address space identifiers (ASIDs) and 
2)~a parallel page table walker. 
TLB accesses from multiple threads to the same page are coalesced.
On a private L1 TLB miss, the shared L2 TLB is probed (\mycirc{4}).
On a shared L2 TLB miss, the page table walker begins a walk,
probing the shared L2 cache
and main memory. 

Figure~\ref{fig:pwcache} compares the performance with multi-address-space 
concurrency of our chosen baseline and the design from \cite{powers-hpca14} against
the ideal scenario where every TLB access is a hit (see
Section~\ref{sec:meth} for our methodology). While Power et al.\ find that a
page walk cache is more effective than a shared L2 TLB~\cite{powers-hpca14},
the design with a shared L2 TLB provides better performance
for all but three workloads, with \emph{13.8\% better performance on average}.
The shared L2 data cache enables a hit rate for page table walks that is
competitive with a dedicated page walk cache, and a shared TLB is a more
effective use of chip area. Hence, we adopt a design with a shared L2 TLB as
the baseline for \titleShort{}. We observe that a 128-entry TLB provides only a
10\% reduction in miss rate over a 64-entry TLB, suggesting that the additional
area needed to double the TLB size is not efficiently utilized.  Thus, we opt for a smaller
64-entry L1 TLB in our baseline. Note that a shared L2 TLB outperforms the page walk
cache for both L1 TLB sizes. Lastly, we find that \emph{both designs incur a
significant performance overhead compared to the ideal case where every TLB
access is a TLB hit.}



\begin{figure}[h]
\centering
\includegraphics[width=\columnwidth]{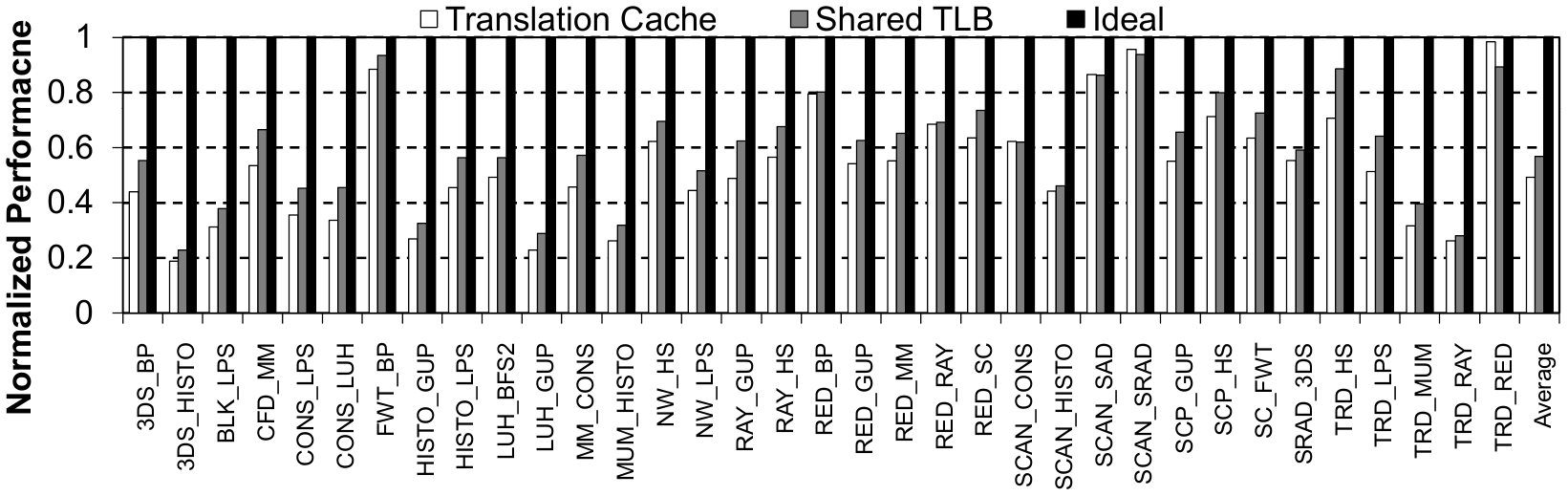}%
\caption{Baseline designs vs. ideal performance.} 
\label{fig:pwcache}
\end{figure}

\section{Design Space Analysis}
\label{sec:motiv}

To inform the design of \titleShort{}, we characterize overheads for address
translation, and consider performance challenges induced by the introduction of
multi-address-space concurrency and contention. 

\subsection{Address Translation Overheads}
\label{sec:tlb-bottleneck}

GPU throughput relies on \emph{fine-grained
multithreading}~\cite{cdc6600,smith-hep} to hide memory latency.  However, we
observe a fundamental tension between address translation and fine-grained
multithreading. The need to cache address translations at a page granularity, combined
with application-level spatial locality, increases the likelihood that
translations fetched in response to a TLB miss will be needed by more than one
thread. Even with the massive levels of parallelism supported by GPUs, we
observe that a small number of outstanding TLB misses can result in the thread
scheduler not having enough ready threads to schedule, which in turn limits the
GPU's most essential latency-hiding mechanism. 

Figure~\ref{fig:stall-tlb} illustrates a scenario where all warps of an application 
access memory. Each box represents a memory instruction, labeled with the issuing warp.
Figure~\ref{fig:stall-tlb-no-translation} shows how the GPU behaves when 
no virtual-to-physical address translation is required. When Warp~A executes a high-latency 
memory access, the core does not stall as long as other warps have schedulable 
instructions: in this case, the GPU core selects from among 
the remaining warps (Warps~B--H) during the next cycle (\mycirc{1}), and continues 
issuing instructions until all requests to DRAM have been sent. 
Figure~\ref{fig:stall-tlb-translation} considers the same scenario \emph{when address translation
is required}. Warp~A misses in the TLB (indicated in red), and stalls 
until the translation is fetched from memory. If threads belonging to Warps~B--D
access data from the same page as the one requested by Warp~A, these warps stall as well (shown in light red)
and perform no useful work (\mycirc{2}). If a TLB miss from Warp~E similarly stalls
Warps~E--G (\mycirc{3}), only Warp~H executes an actual data access (\mycirc{4}). Two phenomena 
harm performance in this scenario. First, warps stalled on TLB misses reduce the availability
of schedulable warps, lowering utilization. Second, TLB miss requests must complete 
before actual the data requests can issue, which reduces the ability of the GPU 
to hide latency by keeping multiple memory requests in flight. 

\begin{figure}[h]
\centering
\subfloat[No virtual-to-physical address translation on critical path]{{
	\includegraphics[width=0.9\columnwidth]{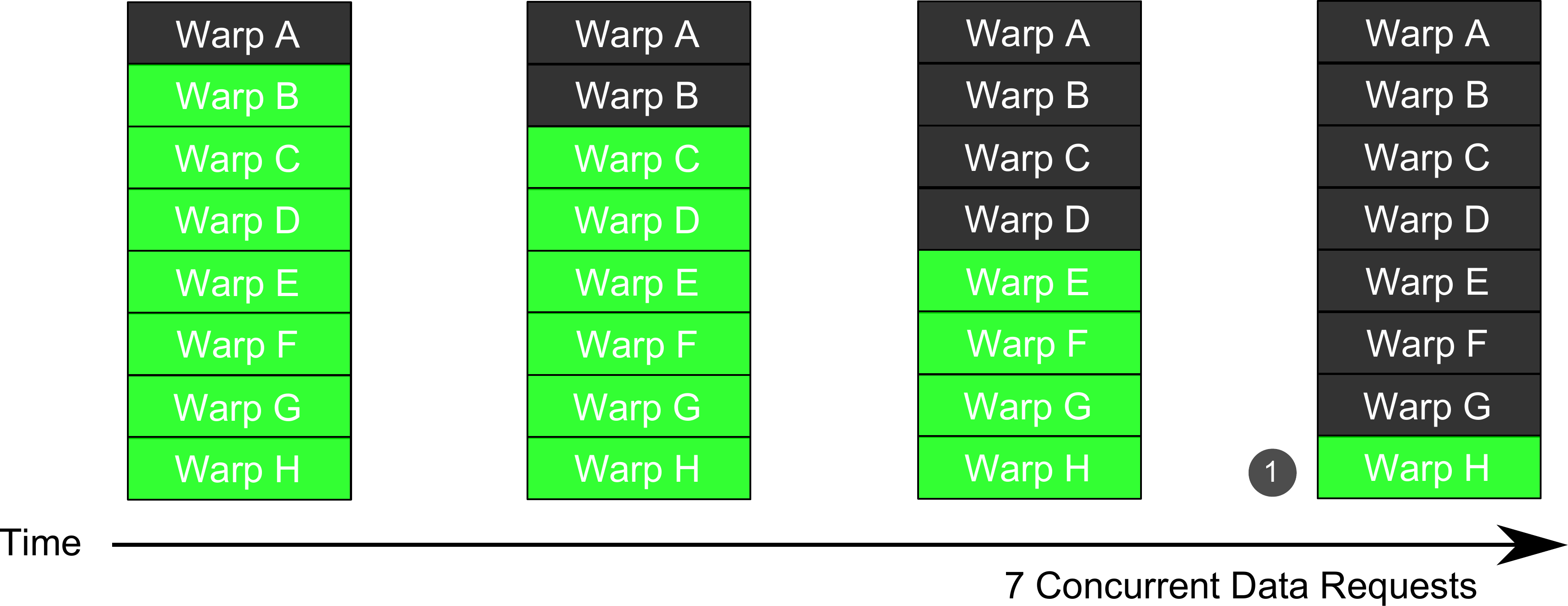}%
	\label{fig:stall-tlb-no-translation}
}}
\qquad
\subfloat[Virtual-to-physical address translation on the critical path]{{
	\includegraphics[width=0.9\columnwidth]{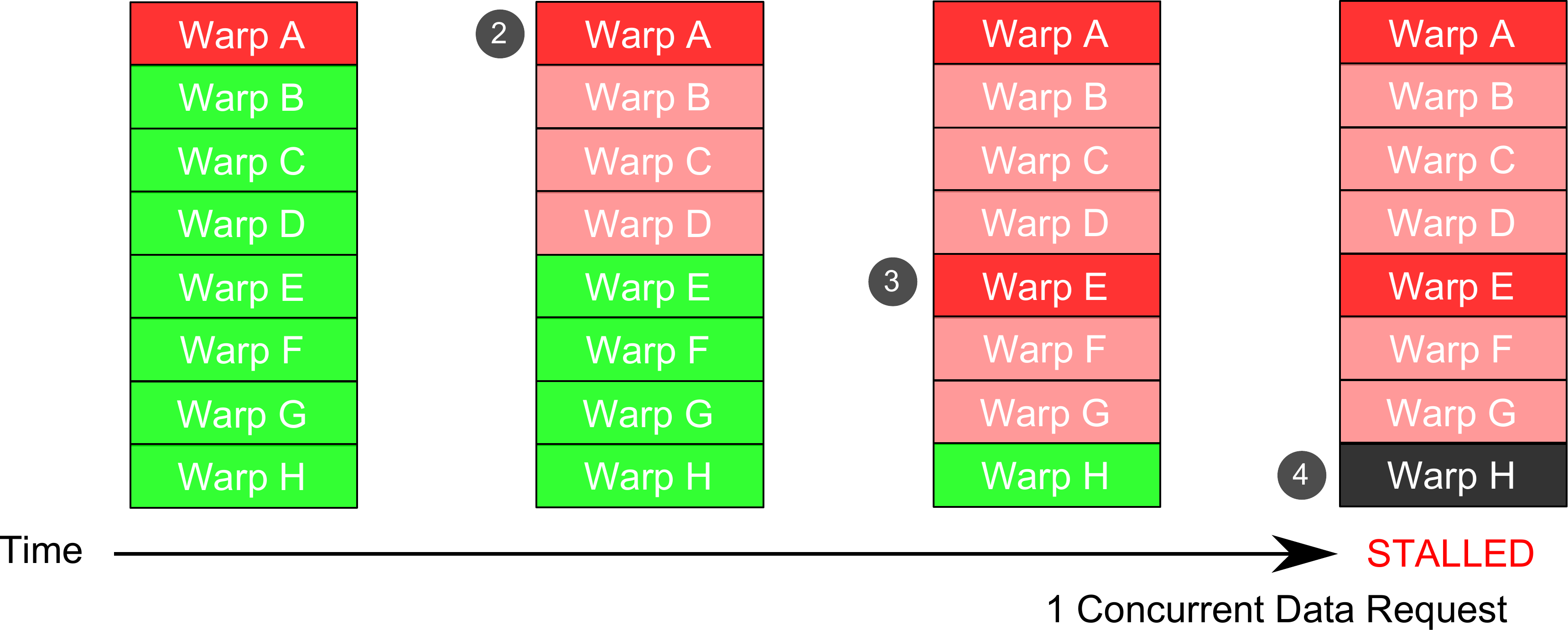}%
	\label{fig:stall-tlb-translation}
}}
\caption{Example bottlenecks created by TLB misses.} 
\label{fig:stall-tlb}
\end{figure}

Figure~\ref{fig:schedulable-warp} shows the number of stalled warps per active
TLB miss, and the average number of maximum concurrent page table walks
(sampled every 10K cycles for a range of applications). In the worst case, a
single TLB miss stalls over 30~warps, and over 50~outstanding TLB misses
contend for access to address translation structures. The large number of
concurrent misses stall a large number of warps, which must wait before issuing
DRAM requests, so minimizing TLB misses and page table walk latency is
critical.

\begin{figure}[h!]
\centering
\includegraphics[width=\columnwidth]{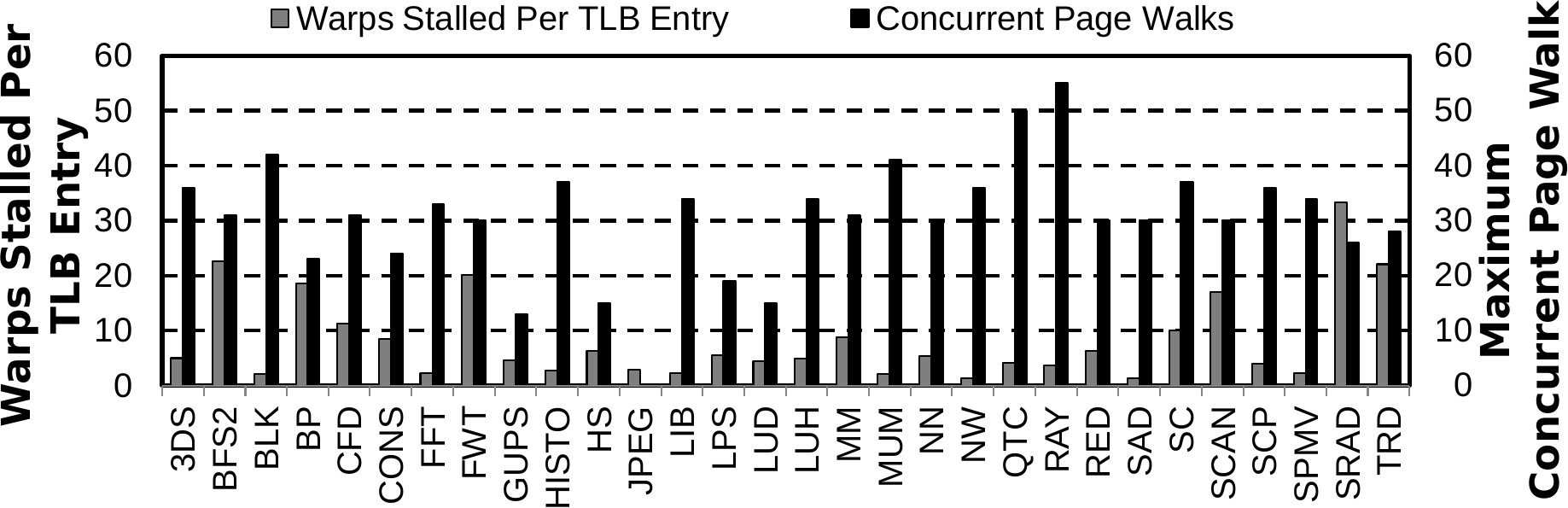}%
\caption{Average number of stalled warp per active TLB miss and number of concurrent page walks. 
}
\label{fig:schedulable-warp}
\end{figure}

\para{Impact of Large Pages.} Larger page size can significantly improve
the coverage of the TLB. However, previous work has observed that the use of large pages
significantly increases the overhead of demand paging in
GPUs~\cite{tianhao-hpca16}. We evaluate this overhead with 2MB page size and find that it results in an average slowdown of 93\%.

\subsection{Interference Induced by Sharing}
\label{sec:motiv-inter-thrashing}

To understand the impact of inter-address-space interference through the memory
hierarchy, we concurrently run two applications using the methodology described
in Section~\ref{sec:meth}. Figure~\ref{fig:tlb-miss-breakdown} shows the TLB
miss breakdown across all workloads: most applications incur significant L1 and
L2 TLB misses. Figure~\ref{fig:interference-real} compares the TLB miss rate
for applications running in isolation to the miss rates under contention.  The
data show that inter-address-space interference through additional thrashing
has a first-order performance impact.

\begin{figure}[h]
\centering
\includegraphics[width=\columnwidth]{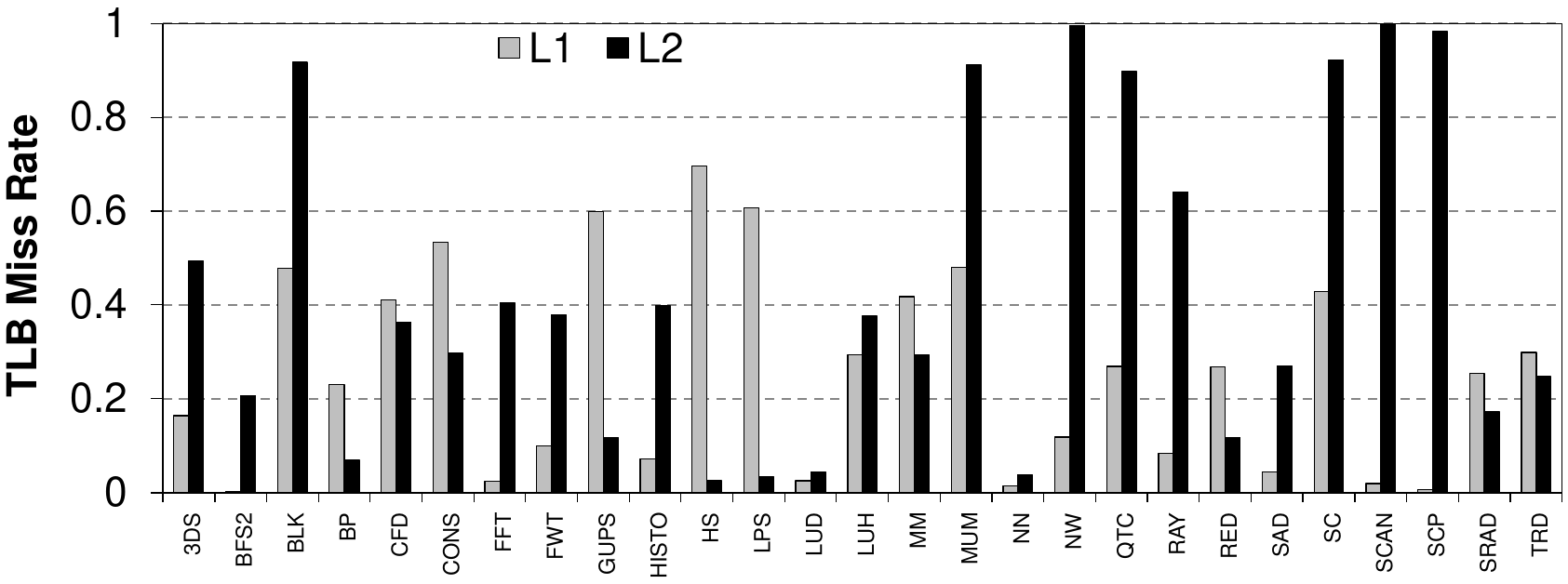}%
\caption{TLB miss breakdown for all workloads.} 
\label{fig:tlb-miss-breakdown}
\end{figure}

\begin{figure}[h]
\centering
\includegraphics[width=\columnwidth]{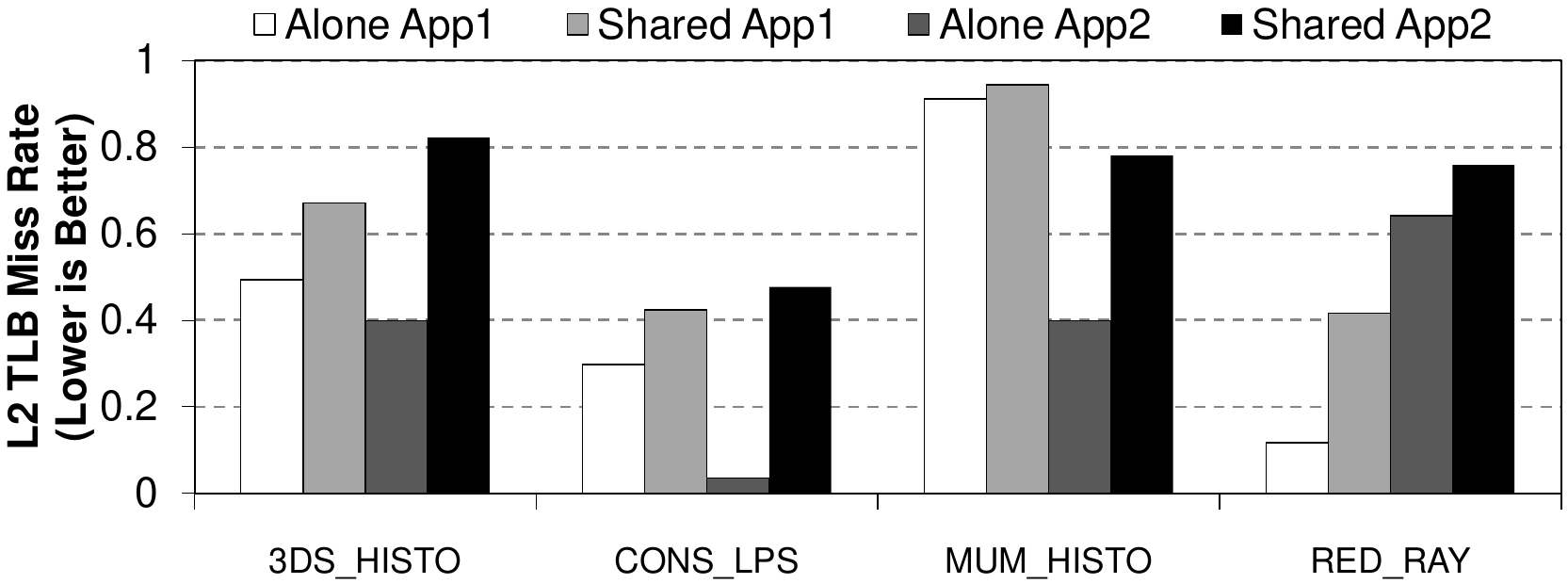}
\caption{Cross-address-space interference in real applications. Each set of
		bars corresponds to a pair of co-scheduled applications, e.g. ``3DS\_HISTO'' 
		denotes the 3DS and HISTO benchmarks running concurrently.} 
\label{fig:interference-real}
\end{figure}

Figure~\ref{fig:interference} illustrates TLB misses in a scenario where
two applications (green and blue) share the GPU. 
In Figure~\ref{fig:interference-a}, the green application issues five parallel TLB
requests, causing the premature eviction of translations for the blue application,
increasing its TLB miss rate (Figure~\ref{fig:interference-b}). The use of a shared 
L2 TLB to cache entries for each application's (non-overlapping) page tables dramatically reduces 
TLB reach. The resulting inter-core TLB thrashing 
hurts performance, and can lead to unfairness and starvation when applications generate 
TLB misses at different rates.  
Our findings of severe performance penalties for increased TLB misses corroborate
previous work on GPU memory designs~\cite{powers-hpca14, pichai-asplos14, abhishek-ispass16}. However,
interference across address spaces can inflate miss rates in ways not addressed by these works,
and which are best managed with mechanisms that are aware of concurrency (as we show in Section~\ref{sec:eval-multi}).~\cjr{we do show this, right?}

\begin{figure}[h]
\centering
\resizebox{\columnwidth}{!}{%
\subfloat[Parallel requests]{{
	\includegraphics[height=80pt]{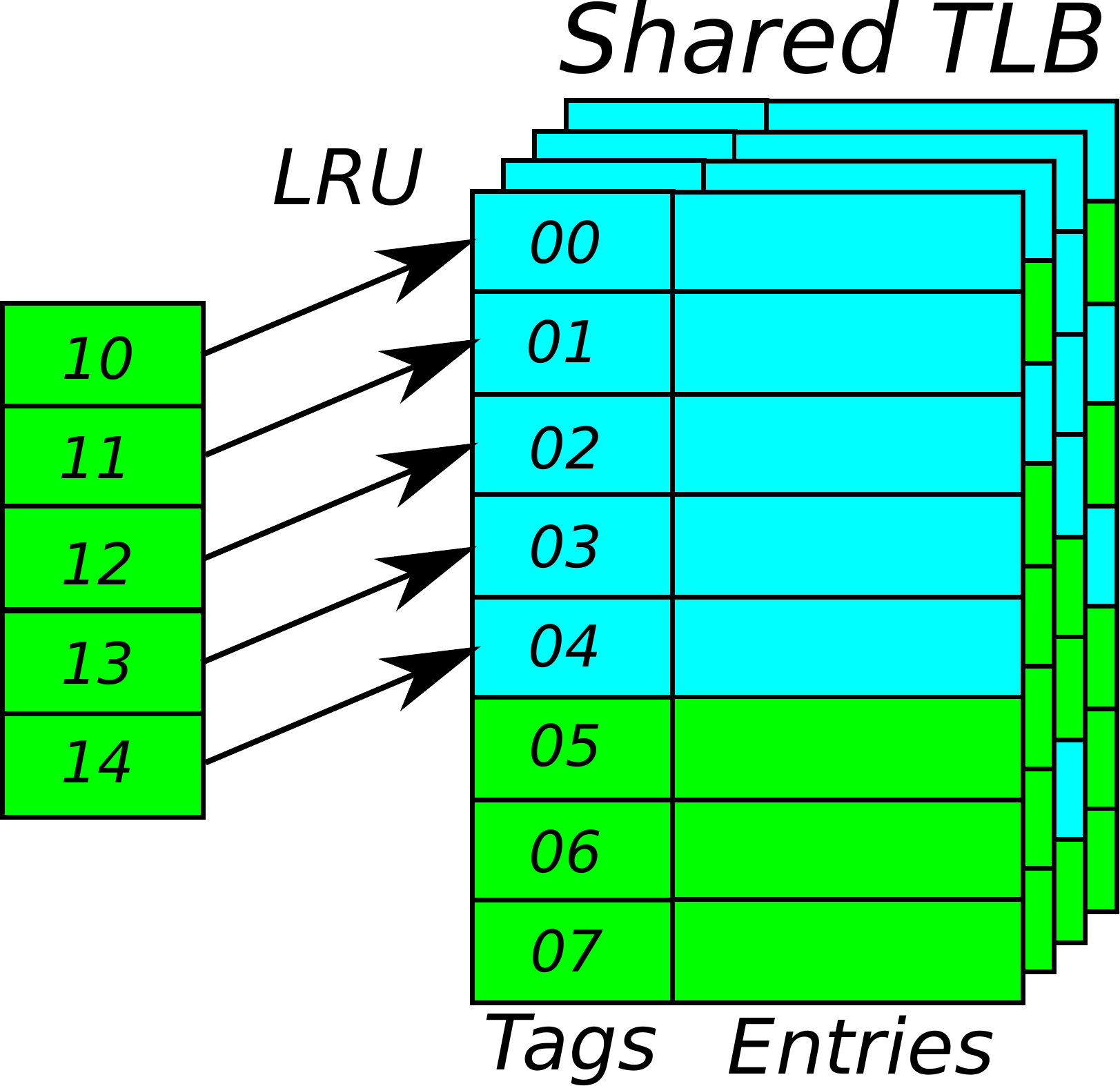}%
	\label{fig:interference-a}
}}%
\subfloat[Conflict]{{
	\includegraphics[height=80pt]{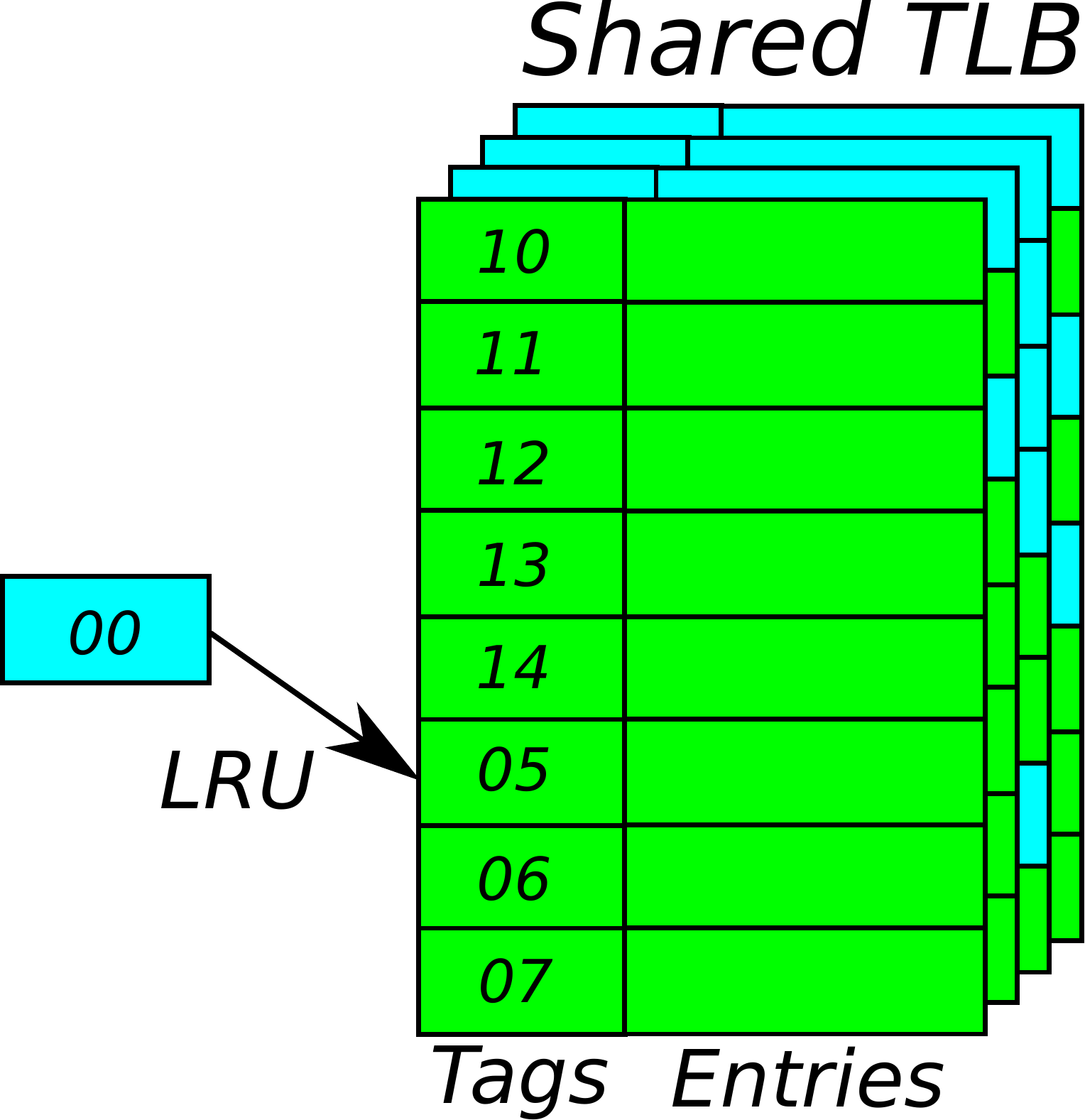}%
	\label{fig:interference-b}
}}%
\qquad
\subfloat[Final state]{{
	\includegraphics[height=80pt]{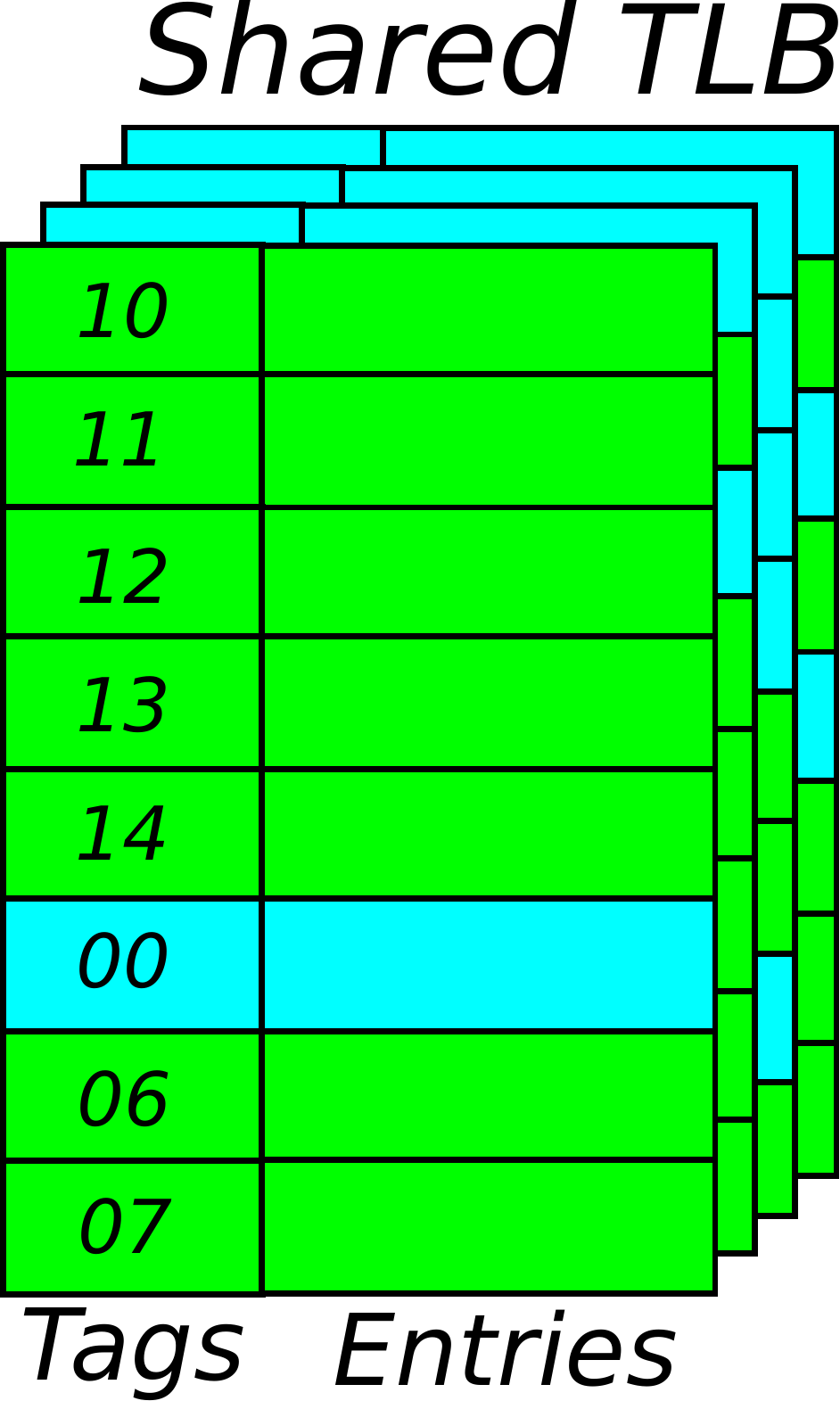}%
	\label{fig:interference-c}
}}%
}
\caption{Cross-address-space TLB interference. \vm{Could this graph use more explanation in the text?}} 
\label{fig:interference}
\end{figure}

\subsection{Interference from Address Translation}
\label{sec:bypassl2cache-motiv}
\label{sec:dram-interference}

\para{Interference at the Shared Data Cache.} 
Prior work~\cite{medic} demonstrated that while cache hits in GPUs reduce the consumption
of off-chip memory bandwidth, cache hits result in a lower load/store 
instruction latency only when \emph{every thread in the warp} hits in the cache. 
In contrast, when a page table walk hits in the shared L2
cache, the cache hit has the potential to help reduce the latency of \emph{other warps}
that have threads which access the same page in memory.  While this makes
it desirable to allow the data generated by the page table walk to consume entries in the shared
cache, TLB-related data can still interfere with and thrash normal data cache
entries, which hurts the overall performance.

Hence, a trade-off exists between prioritizing TLB 
related requests or normal data requests in the GPU memory hierarchy.
Figure~\ref{fig:tlb-level-hit} shows that entries for translation data from
levels closer to the page table root are more likely to be shared across warps, 
and will typically be served by cache hits. Allowing shared structures to cache
page walk data from only the levels closer to the root could alleviate the interference between
low-hit-rate translation data and application data.



\begin{figure}[t]
\centering
\includegraphics[width=\columnwidth]{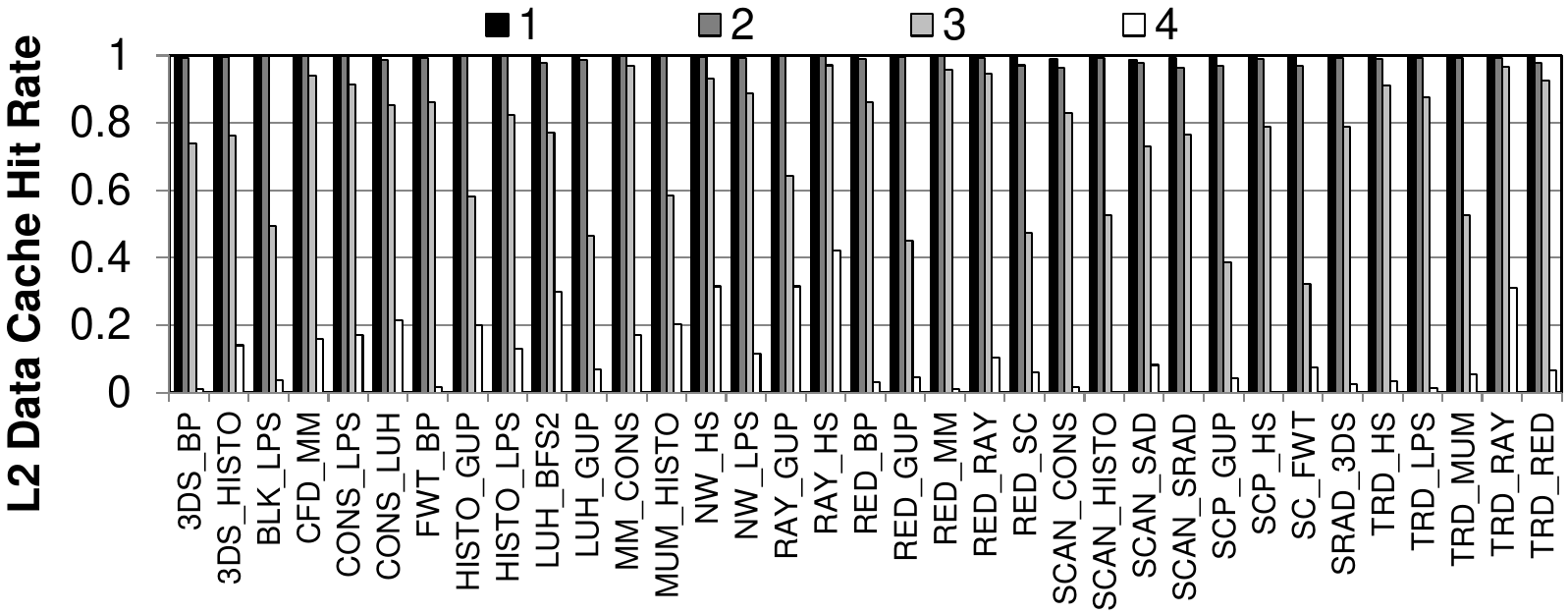}%
\caption{L2 cache hit rate for page walk requests.} 
\label{fig:tlb-level-hit}
\end{figure}

\para{Interference at Main Memory.}
Figure~\ref{fig:dram-util} characterizes the DRAM bandwidth utilization,
broken down between data and address translation requests for applications
sharing the GPU concurrently pairwise.
Figure~\ref{fig:dram-latency} compares the average latency for data requests and 
translation requests. We see that even though page walk requests consume
only 13.8\% of the utilized DRAM bandwidth (2.4\% of the maximum bandwidth),
their DRAM latency is higher than that of data requests, which is particularly egregious since 
data requests that lead to TLB misses stall while waiting for page walks to complete.
The phenomenon is caused by FR-FCFS memory schedulers~\cite{fr-fcfs,frfcfs-patent}, which prioritize accesses that hit in the
row buffer. Data requests from GPGPU applications generally have very
high row buffer locality~\cite{sms,demystify,nmnl-pact13,complexity}, so a scheduler that cannot distinguish 
page walk requests effectively de-prioritizes them, increasing their latency.

\begin{figure}[h]
\centering
\includegraphics[width=\columnwidth]{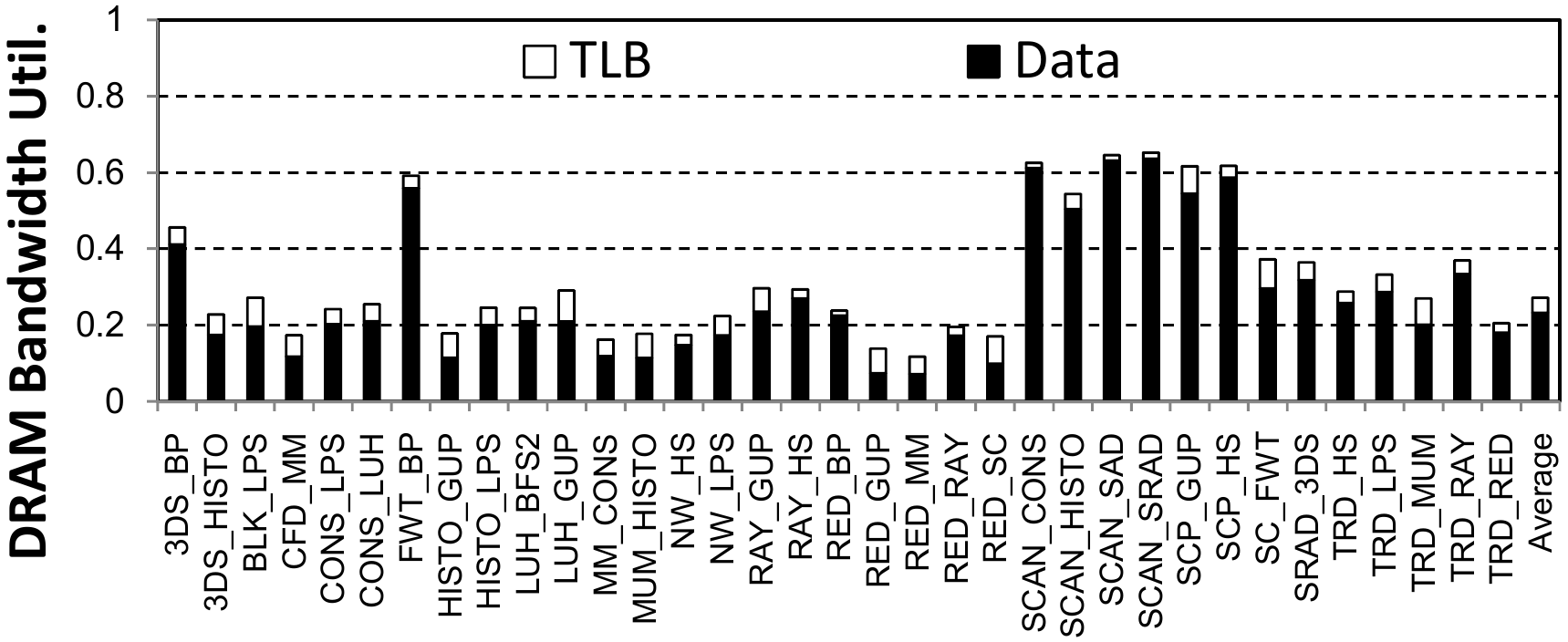}%
\caption{Bandwidth breakdown of two applications.}
\label{fig:dram-util}
\end{figure}

\begin{figure}[h]
\centering
\includegraphics[width=\columnwidth]{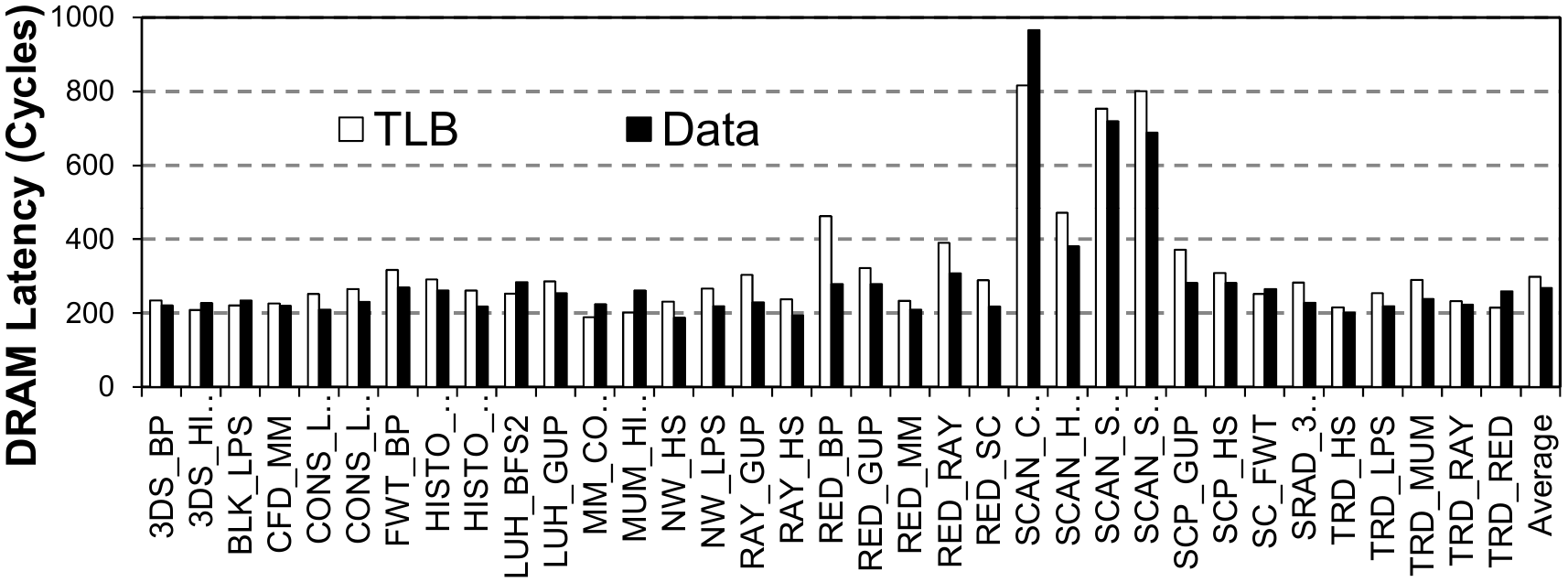}%
\caption{Latency breakdown of two applications.} 
\label{fig:dram-latency}
\end{figure}




\begin{figure*}[ht!]
\centering
\includegraphics[width=1.9\columnwidth]{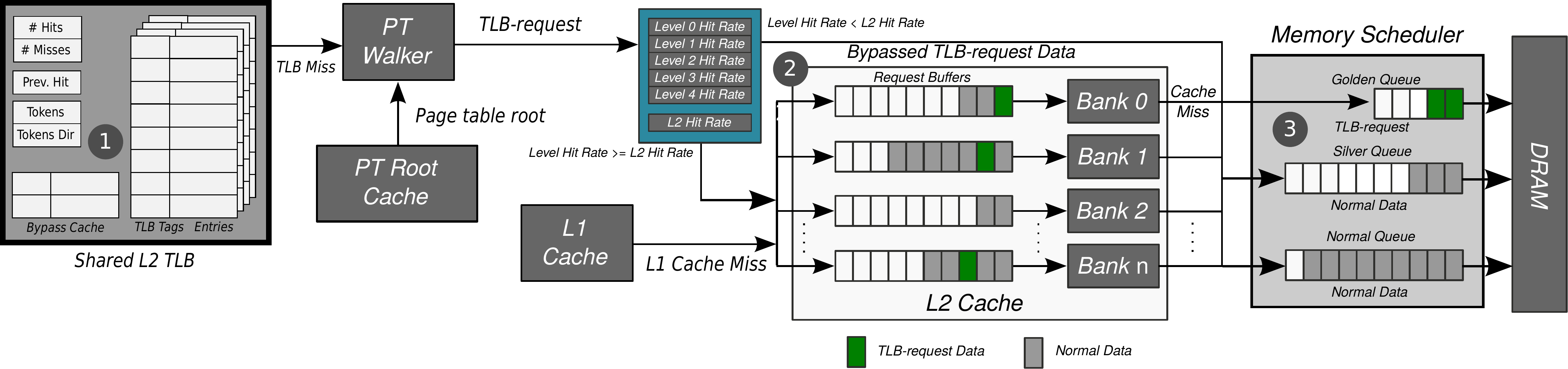}%
\caption{\titleShort design overview.~\cjr{We should expand this to include our page-table root and 
		root caches, please. Also, ``Page-Walk-Related Data'' is misspelled in the graphic.''}} 
\label{fig:overall-design}
\end{figure*}

In summary, we make two important observations about address translation
in GPUs. First, address translation competes with 
the GPU's ability to hide latency through thread-level parallelism, when multiple warps stall on the
TLB misses for a single translation. Second, the GPU's memory-level parallelism
generates interference across address spaces, and between
TLB requests and data requests, which can lead to unfairness and 
increased latency. In light of these observations, \emph{the goal of this work} is to design
mechanisms that alleviate the translation overhead
by 1)~increasing the TLB hit rate through reduced TLB thrashing,
2)~decreasing interference between normal data and
TLB requests in the shared L2 cache, 3)~decreasing TLB miss
latency by prioritizing TLB-related requests in DRAM, and 4)~enhancing memory scheduling to 
provide fairness
without sacrificing DRAM bandwidth utilization.

\section{Design of MASK}
\label{sec:design}

We now introduce \titleLong (\titleShort), a new cooperative resource management
framework and TLB design for GPUs.
Figure~\ref{fig:overall-design} provides a design overview of \titleShort. 
\titleShort employs three components in the memory hierarchy to
reduce address translation overheads while requiring minimal hardware change.
First, we introduce \tlbtokenname
to lower the number of TLB misses and utilize a bypass cache to cache frequently
used TLB entries (\mycirc{1}). Second, we design a \cachebypass mechanism
for TLB requests that significantly increases the shared L2 data cache
utilization, by reducing interference from TLB misses at the shared L2 data cache
(\mycirc{2}). Third, we design an \dramsched to further reduce
interference between TLB requests and data requests from different applications
 (\mycirc{3}). We analyze 
the hardware cost of \titleShort in Section~\ref{sec:overhead}.

\subsection{Memory Protection}

\titleShort uses per-core page table root registers (similar to x86 CR3)
to set the current address space on each core: setting it also sets the value in a 
page table root cache with per-core entries at the L2
layer. The page table root cache is kept coherent with the CR3 value
in the core by draining all in-flight memory requests for
that core when the page table root is set. L2 TLB cache lines are
extended with address-space identifiers (ASIDs); TLB flush operations
target a single shader core, flushing the core's L1 TLB 
and all entries in the L2 TLB with a matching ASID. 

\subsection{Reducing L2 TLB Interference}
\label{sec:fill-bypassing}

Sections~\ref{sec:tlb-bottleneck} and~\ref{sec:motiv-inter-thrashing}
demonstrated the need to minimize TLB miss overheads. \titleShort{} addresses this need 
with a new mechanism called \tlbtokenname. Figure~\ref{fig:tlb-l2-design}a shows 
architectural additions to support \tlbtokenname.
We add two 16-bit counters to track TLB hits and misses per shader core, along
with a small fully-associative bypass cache
to the shared TLB. Figure~\ref{fig:bypass-l2}
illustrates operation of the proposed TLB fill bypassing logic.
When a TLB access arrives (Figure~\ref{fig:bypass-l2-access}), tags for both the shared TLB
(\mycirc{1}) and bypass cache (\mycirc{2}) are probed in parallel. A hit on
either the TLB or the bypass cache yields a TLB hit.

\begin{figure}[h]
\centering
\resizebox{\columnwidth}{!}{%
\subfloat[TLB hit and miss counters]{{%
	\includegraphics[height=80pt]{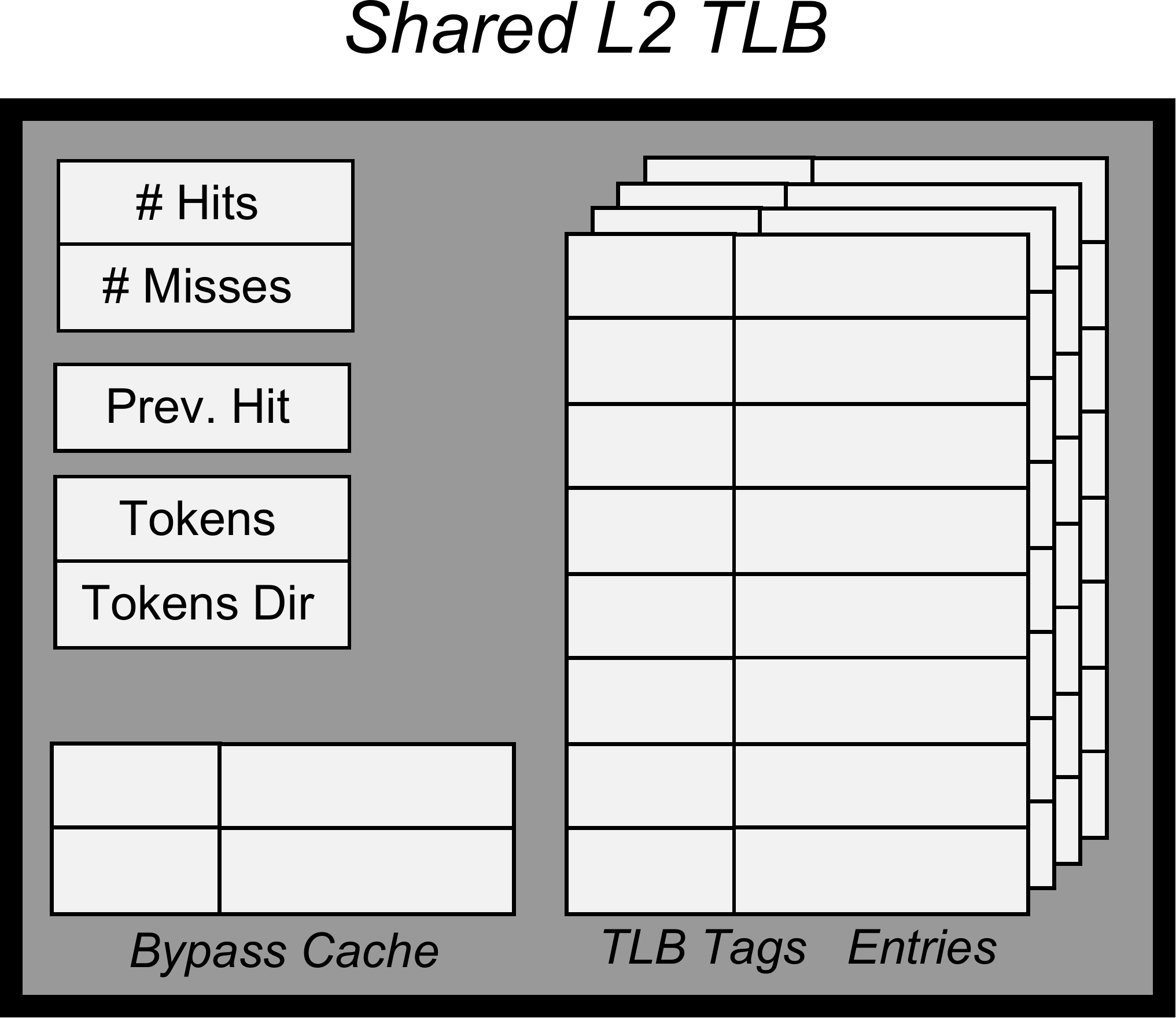}%
	\label{fig:tlb-l2-design-tlb}
}}%
\qquad
\subfloat[TLB counter control logic]{{%
	\includegraphics[height=80pt]{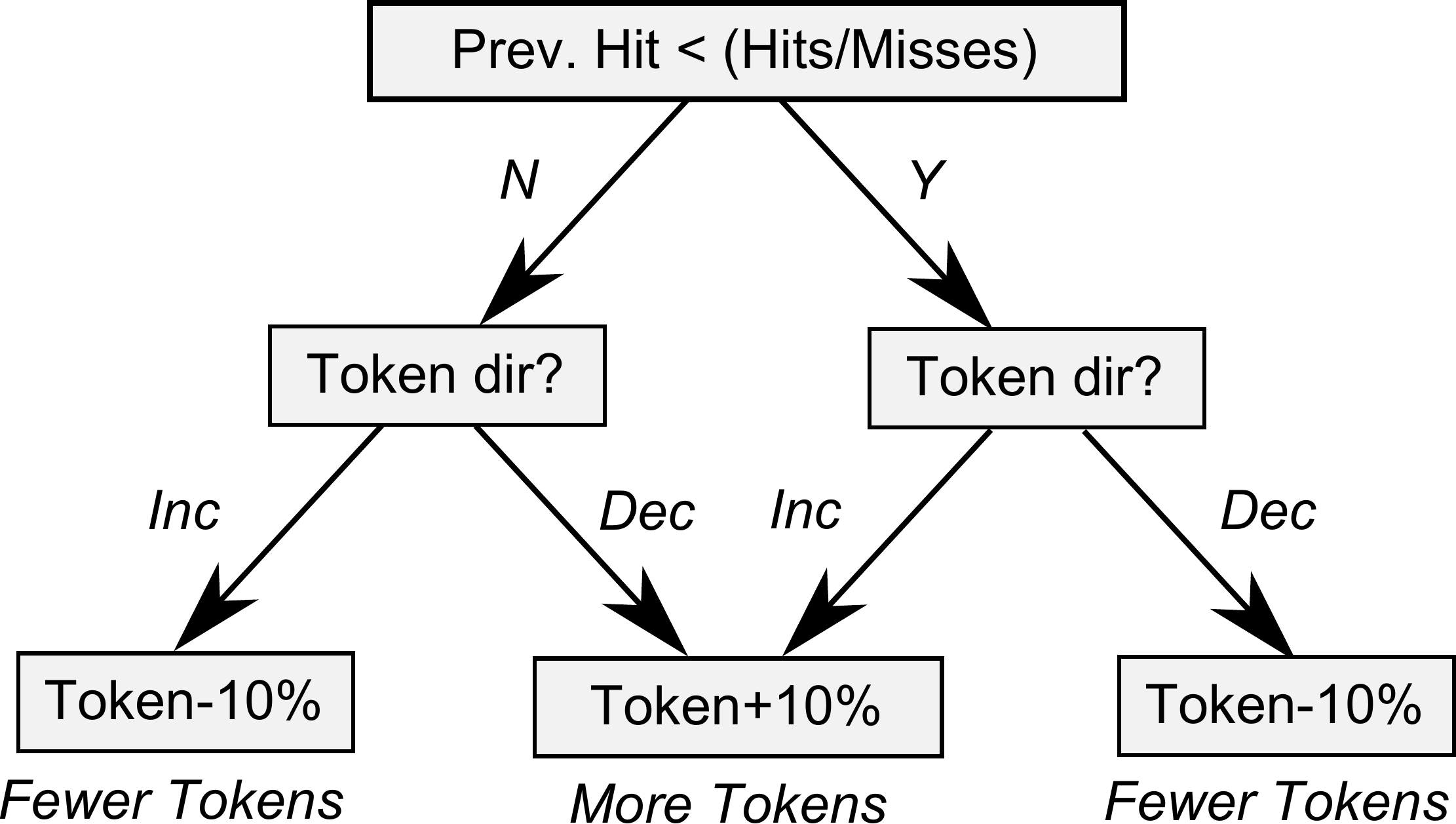}%
	\label{fig:tlb-l2-design-tlb-logic}
}}%
}
\caption{L2 TLB and token assignment logic.}
\label{fig:tlb-l2-design}
\end{figure}

\begin{figure}[h]
\centering
\resizebox{\columnwidth}{!}{%
\subfloat[TLB access]{{%
	\includegraphics[height=80pt]{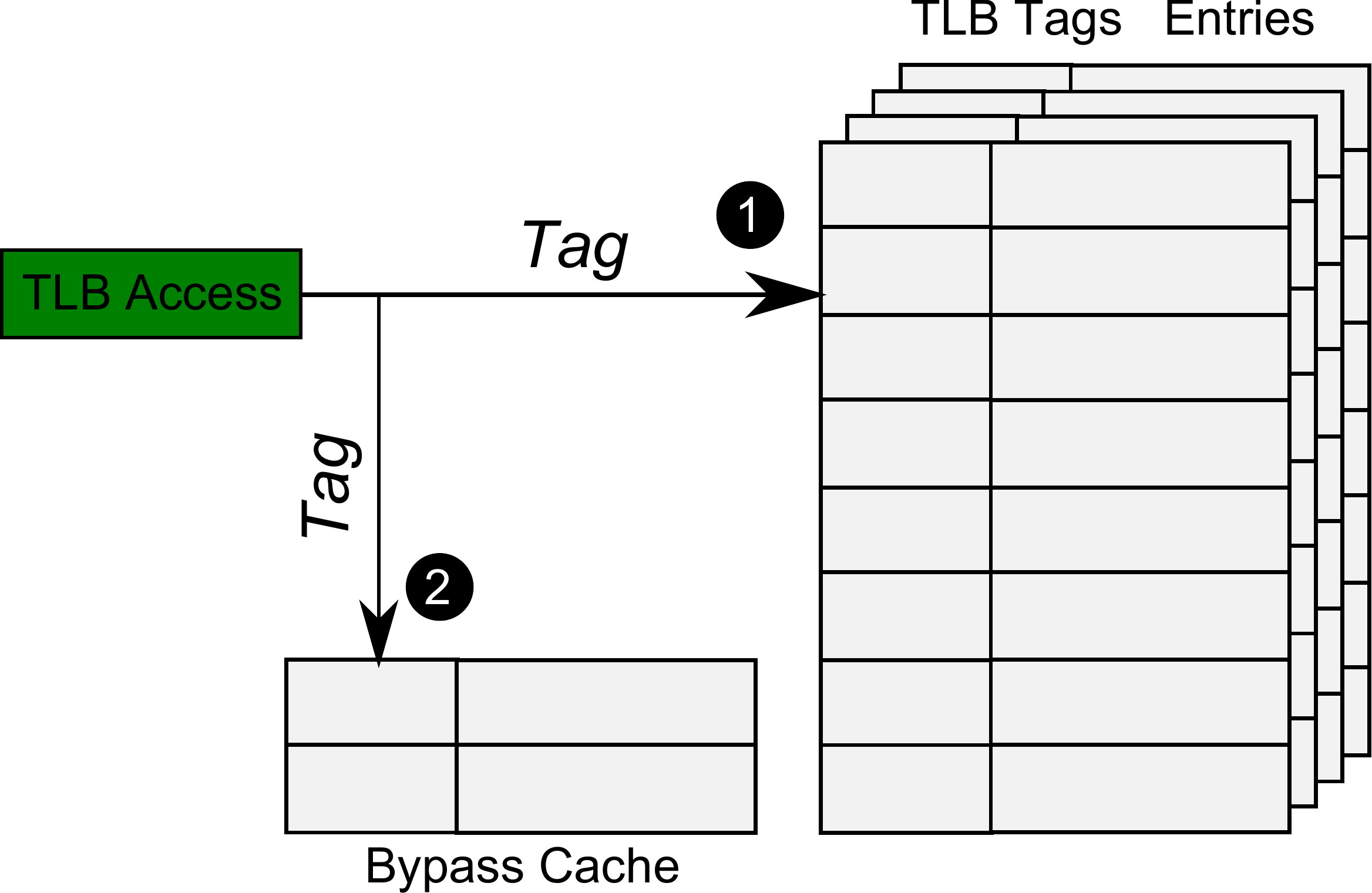}%
	\label{fig:bypass-l2-access}
}}%
\qquad
\subfloat[TLB fill]{{%
	\includegraphics[height=80pt]{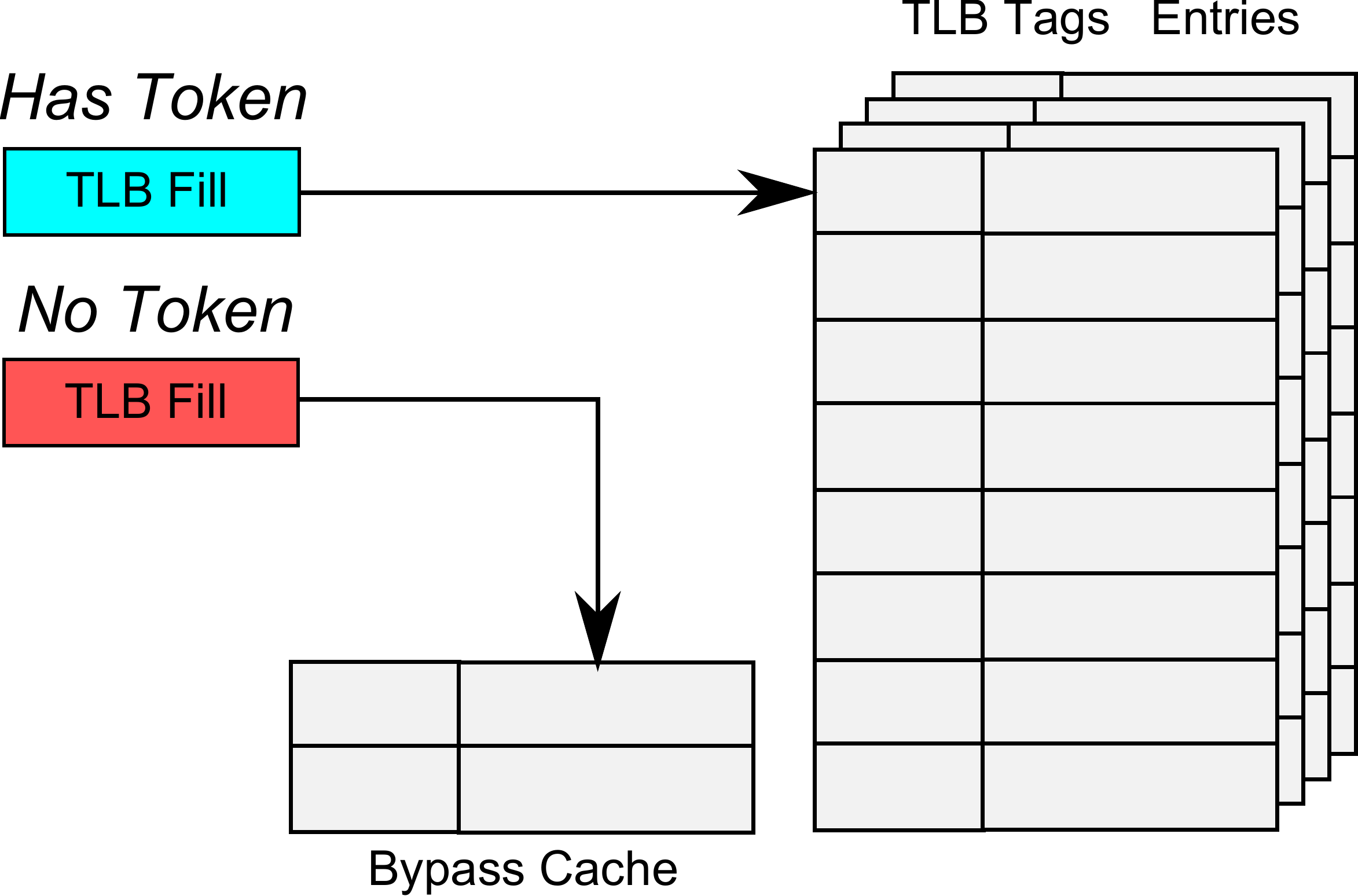}%
	\label{fig:bypass-l2-fill}
}}%
}
\caption{TLB fill bypassing logic in \titleShort{}.} 
\label{fig:bypass-l2}
\end{figure}

To reduce inter-core thrashing at the shared L2 TLB, we use an epoch- and
token-based scheme to limit the number of warps from each shader core that can
fill into (and therefore contend for) the L2 TLB. While every warp can probe
the shared TLB, to prevent thrashing, we allow only warps with tokens to fill into the
shared TLB as shown in Figure~\ref{fig:bypass-l2-fill}. This token-based mechanism requires two components, one to determine
the number of tokens for each application, and one to implement policy for assigning tokens to warps.

\para{Determining the Number of Tokens.} At the beginning of a
kernel, \titleShort performs no bypassing, but tracks the L2 miss rate for each
application and the total number of warps in each core. After the first
epoch,\footnote{\label{foot:epoch}We empirically select an epoch length of 100K cycles.} 
the initial number of tokens ($InitialToken$) is set to a fraction of the
total number of warps per application.
At the end of any subsequent epoch, \titleShort compares the shared L2 TLB miss
rates of the current and previous epoch. If the miss rate decreases or
increases from the previous epoch, \titleShort uses the logic shown in
Figure~\ref{fig:tlb-l2-design-tlb-logic} to decrease or increase the number of tokens
allocated to each application.

\para{Assigning Tokens to Warps.} Empirically, we observe that
1)~warps throughout the GPU cores have mostly even TLB miss rate
distribution; and 2)~it is more beneficial for warps that previously have
tokens to retain their token, as it is more likely that their TLB entries are already
in the shared TLB. We leverage these two observations to
simplify the token assignment logic:  \tlbtokenname simply hands out tokens in round-robin
fashion to all cores in warpID order. The heuristic is effective at reducing thrashing,
as contention at the shared TLB is reduced based on the number of
tokens, and highly-used TLB entries that do not have tokens can still fill
into the bypassed cache.


\para{Bypass Cache.} While \tlbtokenname can reduce
thrashing in the shared TLB, a handful of highly-reused pages from warps with no tokens
may be unable to utilize the shared TLB. To address this, we add a
\emph{bypass cache}, which is a small 32-entry fully associative cache. 
Only warps without tokens can fill the bypass cache.

\para{Replacement Policy.} While it is possible to base the cache
replacement policy on how many warps are stalled per TLB entry and
prioritize TLB entries with more warps sharing an entry, 
we observe small variance across TLB entries on the shared TLB in practice. Consequently,
a replacement policy based on number of warps stalled per
TLB entry actually performs worse than a reuse-based policy. 
Hence, we use LRU replacement policy for L1 TLBs, the shared L2 TLB and the bypass cache.

\subsection{Minimizing Shared L2 Interference}
\label{sec:data-bypassing}

\para{Interference from TLB Requests.} While Power et al.\ propose to coalesce
TLB requests to minimize the cache pressure and performance
impact~\cite{powers-hpca14}, we find that a TLB miss generates shared
cache accesses with varying degrees of locality. Translating addresses through
a multi-level page table (4 levels for \titleShort{}) can generate dependent
memory requests for each level.
This causes significant queuing latency at the shared L2 cache, 
corroborating observations from previous work~\cite{medic}.
Page table entries in levels
closer to the root are more likely to be shared across threads than
entries near the leaves, and more often hit in the shared L2 cache. 

To address both the interference and queuing delay at the shared L2 cache 
we introduce \cachebypass for TLB requests, as shown in 
Figure~\ref{fig:l2-bypass}. To
determine which TLB requests should be bypassed, we leverage our insights from
Section~\ref{sec:bypassl2cache-motiv}.
Because of the sharp drop-off in L2 cache hit rate after the first few
levels, we can simplify the bypassing logic to compare the L2
cache hit rate of each page level for TLB requests
to the L2 cache hit rate for non-TLB requests. We impose L2 cache bypassing
when the hit rate for TLB requests falls below the hit rate for 
non-TLB requests. Memory requests are tagged with three additional 
bits specifying page walk depth, allowing \titleShort{} to differentiate between
request types. These bits are set to
\emph{zero} for normal data requests, and to 7 for any depth higher than
6.\footnote{Note that all experiments done in this paper use a depth of 4.}
~\cjr{Our current page table depth is 4 right?}~\cjr{Don't we need 
separate hit/miss counters for TLB vs non-TLB requests to make this possible?
If so, should fix Figure 12.}

\begin{figure}[h]
\centering
\includegraphics[width=\columnwidth]{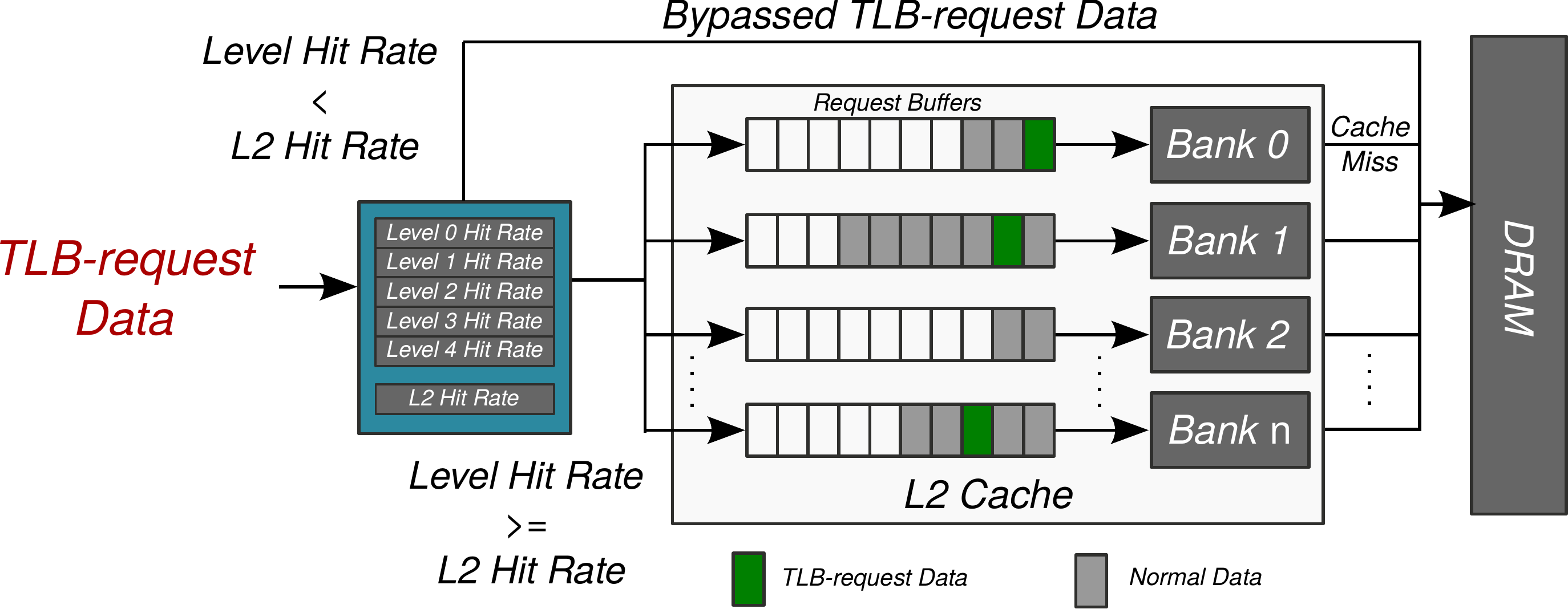}%
\caption{Design of \cachebypass.} 
\label{fig:l2-bypass}
\end{figure}



\subsection{Minimizing Interference at Main Memory}
\label{sec:dram-sched}

Section~\ref{sec:dram-interference} demonstrates two different
types of interference at main memory. Normal data
requests can interfere with TLB requests, and data requests from
multiple applications can interfere with each other. \titleShort{}'s
memory controller design mitigates both forms of interference using 
an \dramsched{}.

\titleShort{}'s \dramsched breaks the traditional DRAM request buffer into three
separate queues, as shown in Figure~\ref{fig:overall-design}. The first queue,
called the \goldQ{}, contains a small FIFO queue.\footnote{We observe
that TLB-related requests have low row locality. Thus, we use a FIFO queue to
further simplify the design.}  TLB-related requests always go to the
\goldQ{}, while non-TLB-related requests go the other two larger
queues (similar to the size of a typical DRAM request buffer size).  The second
queue, called the \silverQ{}, contains data request from \emph{one}
selected application. The last queue, called the \normalQ, contains
data requests from all other applications. The \goldQ is used to prioritize
TLB misses over data requests, while the \silverQ ensures that DRAM bandwidth
is fairly distributed across applications.


Our \dramsched always prioritizes requests in 
the \goldQ over requests in the \silverQ, which are prioritized over requests in the \normalQ. Applications take
turns being assigned to the \silverQ based on two factors: the number of
concurrent page walks, and the number of warps stalled per active TLB miss.
The number of requests each application can add to the \silverQ is
shown in Equation~\ref{eq:silver}.
Application ($App_i$) inserts $thres_i$ requests into the \silverQ.
Then, the next application ($App_{i+1}$) is allowed to send 
$thres_{i+1}$ requests to the \silverQ.  Within each queue,
FR-FCFS~\cite{fr-fcfs,frfcfs-patent} is used to
schedule requests.

\begin{equation}
thres_i = thres_{max} \frac{Concurrent_i * WrpStalled_i}{\sum_{j=1}^{numApp}{Concurrent_j*WrpStalled_j}}
\label{eq:silver}
\end{equation}
\vm{Is there any way to make this equation easier to read? Should it be numbered if it's the only one?}

To track the number of outstanding concurrent page walks, we add a 6-bit counter per
application to the shared TLB.\footnote{We leave techniques to virtualize this counter
for more than 64 applications as future work.} This counter tracks of the \emph{maximum}
number of TLB miss queue, and is used as $Concurrent_i$ in Equation~\ref{eq:silver}. To track
the number of warps stalled per active TLB, we add a 6-bit counter to the TLB
MSHRs, to track the maximum number of warps that hit in
each MSHR entry. This number is used for $WrpStalled_i$. Note that the
\dramsched resets all of these counters every epoch.\cref{foot:epoch}

We find that the number of concurrent TLB requests that go to each
memory channel is small, so our design has an additional benefit of lowering page table
walk latency while minimizing interference.
The \silverQ prevents
bandwidth-heavy applications from interfering with applications utilizing the 
queue, which in turn prevents starvation. It also minimizes the reduction in total
bandwidth utilization, as the per-queue FR-FCFS scheduling policy ensures that
application-level row buffer locality is preserved.~\cjr{maybe float this up earlier 
and replace my attempt to characterize the high level policy goal.}

\subsection{Page Faults and TLB Shootdowns}
\label{sec:mech-other}

Address translation inevitably introduces page faults.  Our design can
be extended to use techniques from previous works, such as performing
copy-on-write for minor faults~\cite{powers-hpca14}, and either exception
support~\cite{igpu} or demand paging techniques~\cite{tianhao-hpca16,pascal,cc-numa-gpu-hpca15}
for major faults. 
We leave this as future work, and do not evaluate these overheads. 

Similarly, TLB shootdowns are required when shader cores change address spaces
and when page tables are updated. We do not envision applications that make
frequent changes to memory mappings, so we expect such events to be rare.~\cjr{Maybe we should state a cost model, even though it won't get exercised unless I fix the scheduler in GPGPU-sim.}
Techniques to reduce TLB shootdown overhead~\cite{unitd,tlb-consistency}
are well-explored and can be applied to \titleShort{}.

\section{Methodology}
\label{sec:meth}

We model Maxwell architecture~\cite{maxwell} cores, TLB fill bypassing, bypass cache, and memory
scheduling mechanisms in \titleShort{} using the MAFIA
framework~\cite{mafia}, which is based on GPGPU-Sim 3.2.2~\cite{gpgpu-sim}. We heavily
modify the simulator to accurately model the behavior of CUDA Unified Virtual
Address~\cite{maxwell,pascal} as described below. Table~\ref{table:config} provides details on our baseline GPU
configuration. In order to show that \titleShort{} works on any
GPU architecture, we also evaluate the performance of \titleShort{} on
a Fermi architecture~\cite{fermi}, which we discuss in Section~\ref{sec:750-eval}.


\begin{table}[h!]
\begin{scriptsize}
\centering
\begin{tabular}{ll}
        \toprule
\textbf{System Overview}           &  30 cores, 64 execution unit per core. 8 memory partitions\\
        \cmidrule(rl){1-2}
\textbf{Shader Core Config}           &  1020 MHz, 9-stage pipeline,\\ & 64 threads per warp, GTO scheduler~\cite{ccws}\\
        \cmidrule(rl){1-2}
\textbf{Private L1 Cache}    &  16KB, 4-way associative, LRU, L1 misses are \\ & coalesced before accessing L2, 1 cycle latency \\
        \cmidrule(rl){1-2} 
\textbf{Shared L2 Cache}   &  2MB total, 16-way associative, LRU, 2 cache banks \\ & 2 interconnect ports per memory partition, 10 cycle latency \\
        \cmidrule(rl){1-2} 
\textbf{Private L1 TLB}    &  64 entries per core, fully associative, LRU, 1 cycle latency \\
        \cmidrule(rl){1-2} 
\textbf{Shared L2 TLB}   &  512 entries total, 16-way associative, LRU, 2 ports \\ & per memory partition (16 ports total), 10 cycle latency \\
        \cmidrule(rl){1-2} 
\textbf{DRAM}   & GDDR5 1674 MHz, 8 channels, 8 banks per rank \\ & FR-FCFS scheduler~\cite{fr-fcfs,frfcfs-patent} burst length 8\\
        \cmidrule(rl){1-2} 
\textbf{Page Table Walker}   & 64 threads shared page table walker, traversing \\ & 4-level page table \\
        \bottomrule
\end{tabular}%
\caption{Configuration of the simulated system.}
\label{table:config}%
\end{scriptsize}%
\end{table}%

\para{TLB and Page Table Walk Model.} We modify the MAFIA
framework to accurately model the TLB designs from~\cite{powers-hpca14} and
the \titleShort{} baseline design. We employ the non-blocking
TLB implementation used in the design from Pichai et al.~\cite{pichai-asplos14}.
Each core has a private L1 TLB. The page table
walker is shared, and admits up to 64 concurrent threads for walks.  The
baseline design for \titleShort{} adds a shared L2 TLB instead of page walk 
caches (see Section~\ref{sec:pwc}),
with a shared L2 TLB in each memory partition.  Both L1 and
L2 TLB entries contain MSHR entries to track in-flight page table walks. On a
TLB miss, a page table walker generates a series
of dependent requests that probe the L2 data cache and main memory as
needed. To correctly model virtual-to-physical address mapping and dependent memory
accesses for multi-level page walks, we collect traces of all virtual addresses referenced
by each application (executing them to completion), enabling us to pre-populate 
disjoint physical address spaces for each application with valid page tables.

\para{Workloads.}
We randomly select 27 applications from the CUDA SDK~\cite{cuda-sdk},
Rodinia~\cite{rodinia}, Parboil~\cite{parboil},
LULESH~\cite{lulesh,lulesh-origin}, and SHOC~\cite{shoc} suites. We classify
these benchmarks based on their L1 and L2 TLB miss rates into one of four
groups. Table~\ref{table:bench} shows the categorization for each benchmark.
For our multi-application results, we randomly select 35 pairs of applications,
avoiding combinations that select applications from the
\emph{lowL1miss-lowL2miss} category, as these applications are relatively
insensitive to memory protection overheads. 
The application that finishes
first is relaunched to keep the SM full and to properly model contention. 

We divide these pairs into three workload categories based on the number of
applications that are from \emph{highL1miss-highL2miss} category. \emph{0 HMR} contains
workload bundles where \emph{none} of the applications in the bundle are from
\emph{highL1miss-highL2miss}.  \emph{1 HMR} contains workloads where
\emph{only one} application in the bundle is from \emph{highL1miss-highL2miss}.
\emph{2 HMR} contains workloads where \emph{both} applications in the
bundle are from \emph{highL1miss-highL2miss}.

\begin{table}[h!]
\begin{scriptsize}
\centering
\begin{tabular}{|l|l|l|}
        \hline
\textbf{L1 TLB Miss} & \textbf{L2 TLB Miss}   & \textbf{Benchmark Name} \\
        \hline
        \hline
Low & Low   & LUD, NN \\
        \hline
Low & High   & BFS2, FFT, HISTO, NW,  \\
    & & QTC, RAY, SAD, SCP \\
        \hline
High & Low & BP, GUP, HS, LPS \\
        \hline
High & High  & 3DS, BLK, CFD, CONS,  \\
& & FWT, LUH, MM, MUM, RED,\\
& & SC, SCAN, SRAD, TRD \\
        \hline
\end{tabular}%
\caption{Categorization of each benchmark\rachata{cite the paper they are from}.}
\label{table:bench}%
\end{scriptsize}%
\end{table}%

\begin{figure*}[ht!!!]
\centering
\includegraphics[width=2\columnwidth]{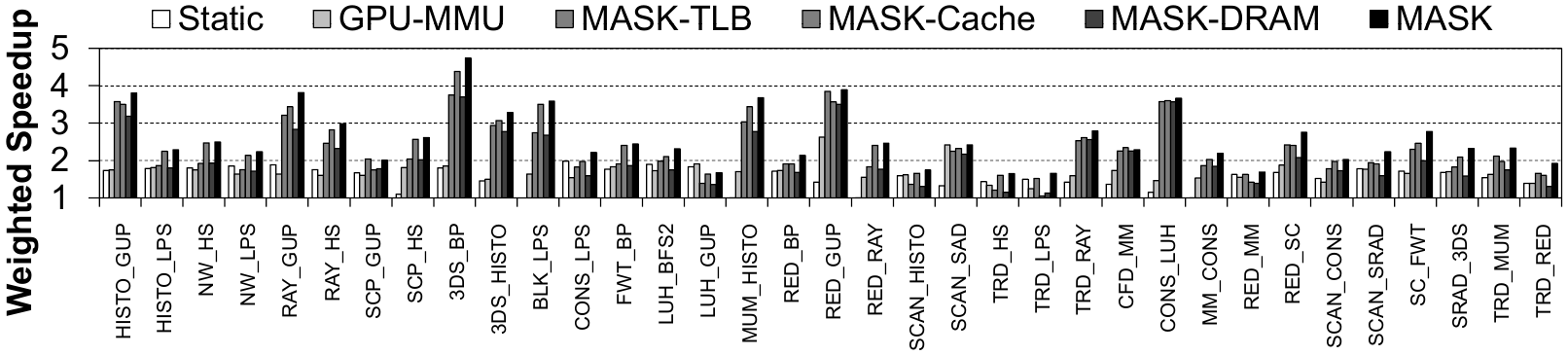}%
\caption{System-wide weighted speedup for multiprogrammed workloads.} 
\label{fig:multi-ws-detail}
\end{figure*}

\para{Evaluation Metrics.} 
We report performance using weighted
speedup~\cite{harmonic_speedup,ws-metric2}, defined as
$\sum{\frac{IPC_{Shared}}{IPC_{Alone}}}$.  $IPC_{alone}$ is the
IPC of an application that runs on the same number of shader cores, but does not
share GPU resources with any other applications, and $IPC_{shared}$ is
the IPC of an application when running concurrently with other
applications. We report the unfairness of each design using
maximum slowdown, defined as $Max{\frac{IPC_{Alone}}{IPC_{Shared}}}$~\cite{ebrahimi-micro09,sms}.

\para{Scheduling and Partitioning of Cores.} The design space for core scheduling
is quite large, and finding optimal algorithms is beyond the scope of this paper. 
To ensure that we model a scheduler that performs reasonably well, we assume an oracle schedule that 
finds the best partition for each pair of applications. For each pair of applications, concurrent 
execution partitions the cores according to the best weighted 
speedup observed for that pair during an exhaustive search over all possible partitionings.


\para{Design Parameters.} \titleShort exposes two configurable parameters:
$InitialTokens$ for \tlbtokenname and $thres_{max}$ for the \dramsched. A sweep
over the range of possible $InitialTokens$ values reveals less than 1\% performance
variance, as \tlbtokenname is effective at reconfiguring the total number
of tokens to a steady-state value (shown in Figure~\ref{fig:bypass-l2}).
In our evaluation, we set $InitialTokens$ to 80\%. We set $thres_{max}$
to $500$ empirically.

\section{Evaluation}

We compare the performance of \titleShort against three designs.  The
first, called \emph{Static}, uses a static spatial partitioning of resources, where an
oracle is used to partition GPU cores, but the shared L2 cache and memory
channels are partitioned equally to each application. This design is intended
to capture key design aspects of NVIDIA GRID and AMD FirePro---however,
insufficient information is publicly available to enable us to build a higher
fidelity model. The second design, called \emph{GPU-MMU}, models the flexible spatial
partitioning GPU MMU design proposed by Power et al.~\cite{powers-hpca14}.\footnote{Note that we
use the design in Figure~\ref{fig:tlb-mask-baseline} instead of the one in
Figure~\ref{fig:tlb-powers-hpca14}, as it provides better performance for the
workloads that we evaluate.} The third design we compare to is an \emph{ideal} scenario,
where every single TLB access is a TLB hit. We also report performance impact for individual
components of \titleShort{}: \tlbtokenname(\titleShort{}-TLB),
\cachebypass(\titleShort{}-Cache), and \dramsched(\titleShort{}-DRAM).

\subsection{Multiprogrammed Performance}
\label{sec:eval-multi}

Figures~\ref{fig:multi-ws} and~\ref{fig:multi-ws-detail} compare the weighted speedup of multiprogrammed
workloads for \titleShort{}, as well as each of the components of
\titleShort{}, against Static and GPU-MMU. Each group of bars in the figure
represents a pair of co-scheduled benchmarks. Compared to GPU-MMU,
\titleShort{} provides 45.2\% additional speedup. We also found that
\titleShort performs only 23\% worse than the ideal scenario where the TLB always
hits. We observe that \titleShort{} provides 43.4\% better aggregate throughput
(system wide IPC) compared to GPU-MMU. Compared to the Static baseline, where
resources are statically partitioned, both GPU-MMU and \titleShort provide
better performance, because when an application stalls for concurrent TLB
misses, it does not use other shared resources such as DRAM bandwidth. During
such stalls, other applications can utilize these resources. When multiple
GPGPU applications run concurrently, TLB misses from two or more applications
can be staggered, increasing the likelihood that there will be heterogeneous
and complementary resource demand.

\begin{figure}[ht!!!]
\centering
\includegraphics[width=\columnwidth]{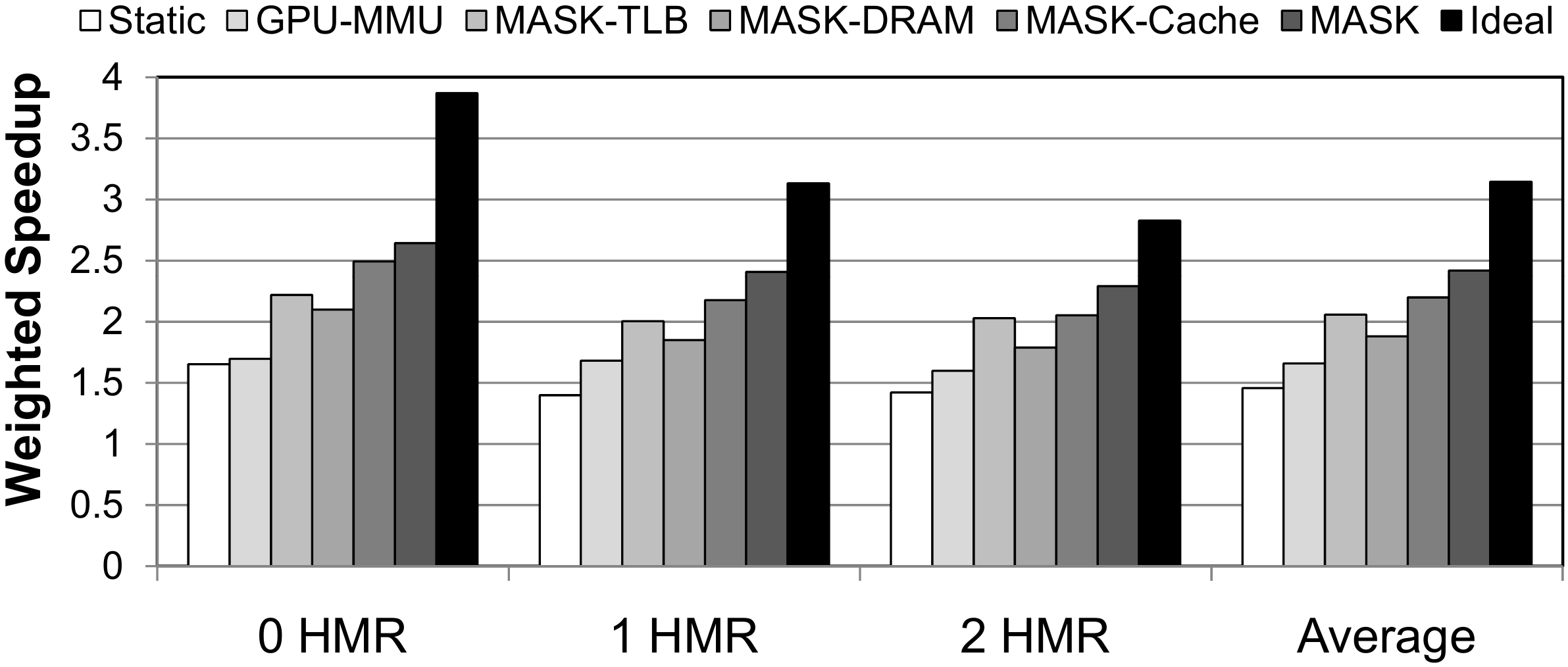}%
\caption{System-wide weighted speedup for multiprogrammed workloads.} 
\label{fig:multi-ws}
\end{figure}

Figure~\ref{fig:multi-unfair} compares unfairness in \titleShort{} against the
GPU-MMU and Static baselines. On average, our mechanism reduces
unfairness by 22.4\% compared to GPU-MMU.  As the number of tokens for each
application changes based on the TLB miss rate, applications that benefit more
from the shared TLB are more likely to get more tokens, causing applications
that do not benefit from shared TLB space to yield that shared TLB space to
other applications. Our application-aware token distribution mechanism and TLB
fill bypassing mechanism can work in tandem to reduce the amount of
inter-application cache thrashing observed in
Section~\ref{sec:motiv-inter-thrashing}. Compared to statically partitioning
resources in Static, allowing both applications access to all of the shared
resources provides better fairness.  On average, \titleShort{} reduces
unfairness by 30.7\%, and a handful of applications benefit by as much as
80.3\%.

\begin{figure}[h]
\centering
\includegraphics[width=1\columnwidth]{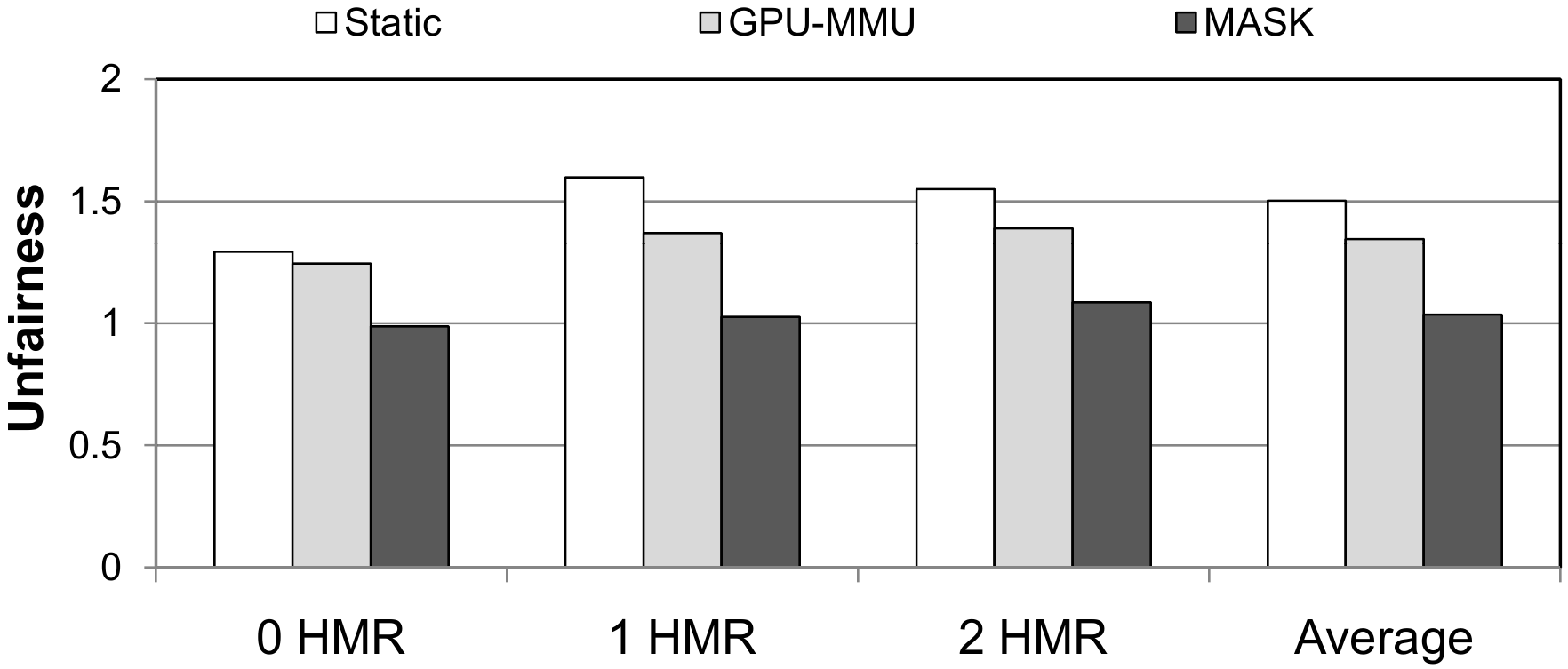}%
\caption{Max unfairness of GPU-MMU and \titleShort.} 
\label{fig:multi-unfair}
\end{figure}

\para{Individual Application Analysis.} \titleShort{} provides better
throughput on all applications sharing the GPU due to reduced TLB miss
rates for each application. The per-application L2 TLB miss rates are reduced
by over 50\% on average, which is in line with the system-wide miss rates observed
in Figure~\ref{fig:tlb-baseline}. Reducing the number of TLB misses through
the TLB fill bypassing policy (Section~\ref{sec:fill-bypassing}), and
reducing the latency of TLB misses through the shared L2 bypassing
(Section~\ref{sec:data-bypassing}) and the TLB- and application-aware DRAM
scheduling policy (Section~\ref{sec:dram-sched}) enables significant
performance improvement.

In some cases, running two applications concurrently provides better speedup
than running the application alone (e.g., RED-BP, RED-RAY, SC-FWT). We
attribute these cases to substantial improvements (more than 10\%) of two
factors: a lower L2 queuing latency for bypassed TLB requests, and a higher L1
hit rate when applications share the L2 and main memory with other
applications.

\subsection{Component-by-Component Analysis}
\label{sec:component-analysis}

\para{Effectiveness of \tlbtokenname.} 
Table~\ref{fig:tlb-baseline} compares the TLB hit rates of GPU-MMU and
\titleShort-TLB. We show only GPU-MMU results for TLB hit rate experiments, as the
TLB hit behavior for Static and GPU-MMU are similar. \titleShort-TLB increases
TLB hit rates by 49.9\% on average, which we attribute to \tlbtokenname. First,
\tlbtokenname reduces the number of warps utilizing the shared TLB entries,
which in turn reduces the miss rate. Second, the bypass cache can store
frequently-used TLB entries that cannot be filled in the traditional TLB.
Table~\ref{fig:tlb-bypass-hit} confirms this, showing the hit rate of the
bypass cache for \titleShort-TLB. From Table~\ref{fig:tlb-baseline}
and Table~\ref{fig:tlb-bypass-hit}, we conclude that the TLB-fill bypassing component
of \titleShort successfully reduces thrashing and ensures that frequently-used
TLB entries stay cached. 



\begin{table}[h!]
\centering
\begin{tabular}{|c|c|c|c|c|}
        \hline
\textbf{Shared TLB} & \textbf{0 HMR}   & \textbf{1 HMR} & \textbf{2 HMR} & \textbf{Average} \\
\textbf{Hit Rate} & & & & \\
        \hline
        \hline
GPU-MMU & 47.8\% & 45.6\% & 55.8\% & 49.3\% \\
        \hline
\titleShort-TLB & 68.1\% & 75.2\% & 76.1\% & 73.9\% \\
        \hline
\end{tabular}%
\caption{Aggregate Shared TLB hit rates.} 
\label{fig:tlb-baseline}
\end{table}%


\begin{table}[h!]
\centering
\begin{tabular}{|c|c|c|c|c|}
        \hline
\textbf{Bypass Cache} & \textbf{0 HMR}   & \textbf{1 HMR} & \textbf{2 HMR} & \textbf{Average} \\
\textbf{Hit Rate} & & & & \\
        \hline
        \hline
\titleShort-TLB & 63.9\% & 66.6\% & 68.8\% & 66.7\% \\
        \hline
\end{tabular}%
\caption{TLB hit rate for bypassed cache.}
\label{fig:tlb-bypass-hit}%
\end{table}%

\para{Effectiveness of \cachebypass.}
Table~\ref{fig:tlb-data-hit} shows the average L2 data cache hit rate for
TLB requests. For requests that fill into the shared L2 data cache, \cachebypass is
effective in selecting which blocks to cache, resulting in a TLB request hit rate
that is higher than 99\% for all of our workloads.
At the same time, \cachebypass minimizes the
impact of bypassed TLB requests, leading to 17.6\% better
performance on average compared to GPU-MMU, as shown in Figure~\ref{fig:multi-ws}.


\begin{table}[h!]
\centering
\begin{tabular}{|c|c|c|c|c|}
        \hline
\textbf{L2 Data Cache} & \textbf{0 HMR}   & \textbf{1 HMR} & \textbf{2 HMR} & \textbf{Average} \\
\textbf{Hit Rate} & & & & \\
        \hline
        \hline
GPU-MMU & 71.7\% & 71.6\% & 68.7\% & 70.7\% \\
        \hline
\titleShort-Cache & 97.9\% & 98.1\% & 98.8\% & 98.3\% \\
        \hline
\end{tabular}%
\caption{L2 data cache hit rate for TLB requests.\cjr{change Y axis label, which is correct, but invites confusion!}} 
\label{fig:tlb-data-hit}
\end{table}%

\rachata{See if there are any changes in L2 nonTLB hit rate. Might improve due to lower contention.}

\para{Effectiveness of \dramsched.}
While the impact of the DRAM scheduler we propose is minimal for many applications,
(the average improvement across \emph{all} workloads is just 0.83\% in
Figure~\ref{fig:multi-ws}), we observe that a few applications that suffered more severely from
interference (see Figures~\ref{fig:dram-util}
and~\ref{fig:dram-latency})
can significantly benefit from our scheduler, since it prioritizes TLB-related
requests. Figures~\ref{fig:multi-dram-util}
and~\ref{fig:multi-dram-lat} compare the DRAM bandwidth utilization and
DRAM latency of GPU-MMU and \titleShort-DRAM for workloads that benefit
from \dramsched.  When our DRAM scheduler policy is employed,
SRAD from the SCAN-SRAD pair sees a 18.7\% performance improvement, while both
SCAN and CONS from SCAN-CONS have performance gains of 8.9\% and 30.2\%, 
respectively. In cases where the DRAM latency is high, the DRAM scheduler policy 
reduces the latency of TLB requests by up to 10.6\% (SCAN-SAD),
while increasing DRAM bandwidth utilization by up to 5.6\%
(SCAN-HISTO). 

\begin{figure}[h]
\centering
\resizebox{\columnwidth}{!}{%
\subfloat[DRAM Bandwidth Utilization]{{
	\includegraphics[height=80pt]{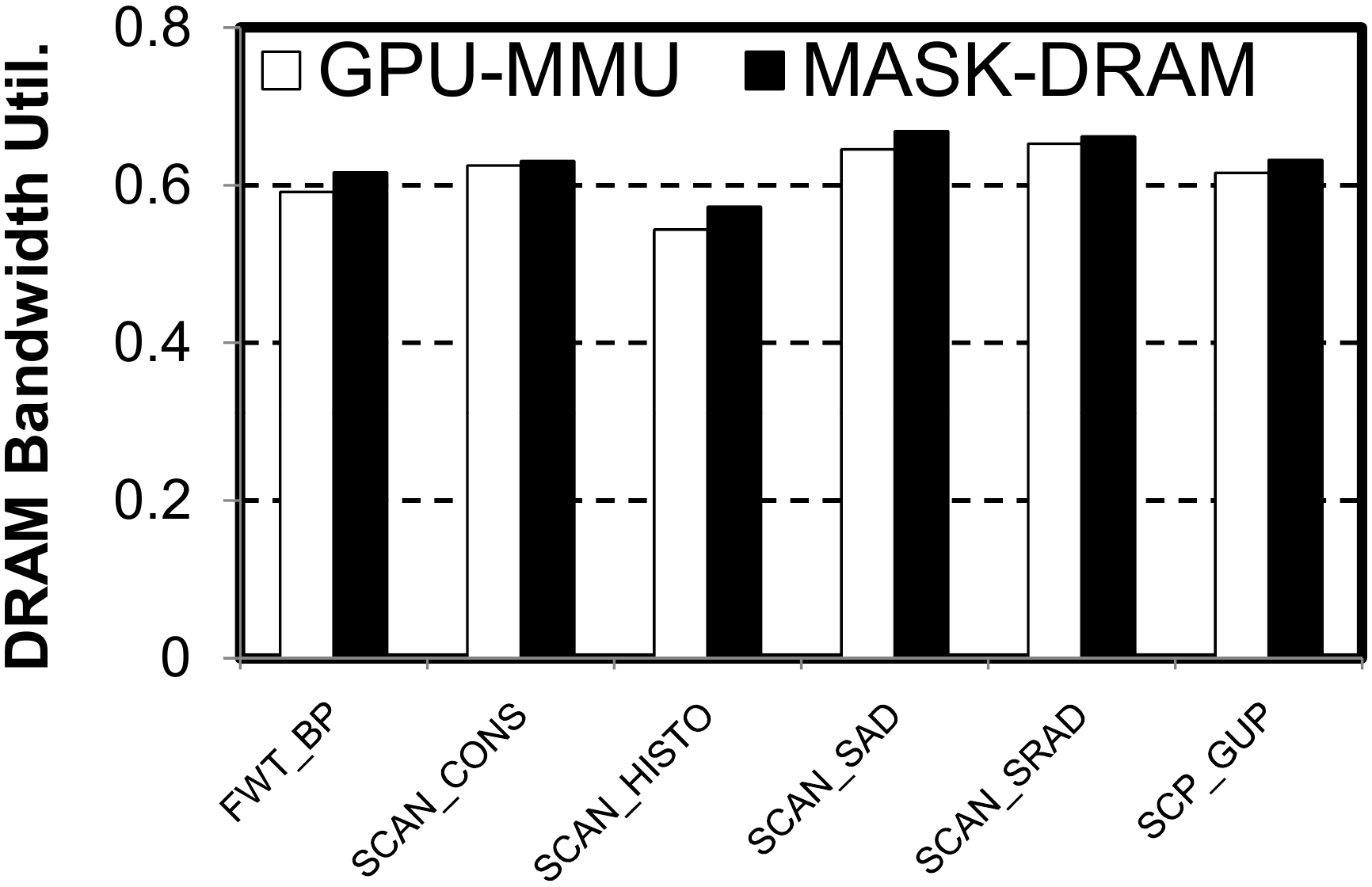}%
	\label{fig:multi-dram-util}
}}%
\subfloat[DRAM Latency]{{
	\includegraphics[height=80pt]{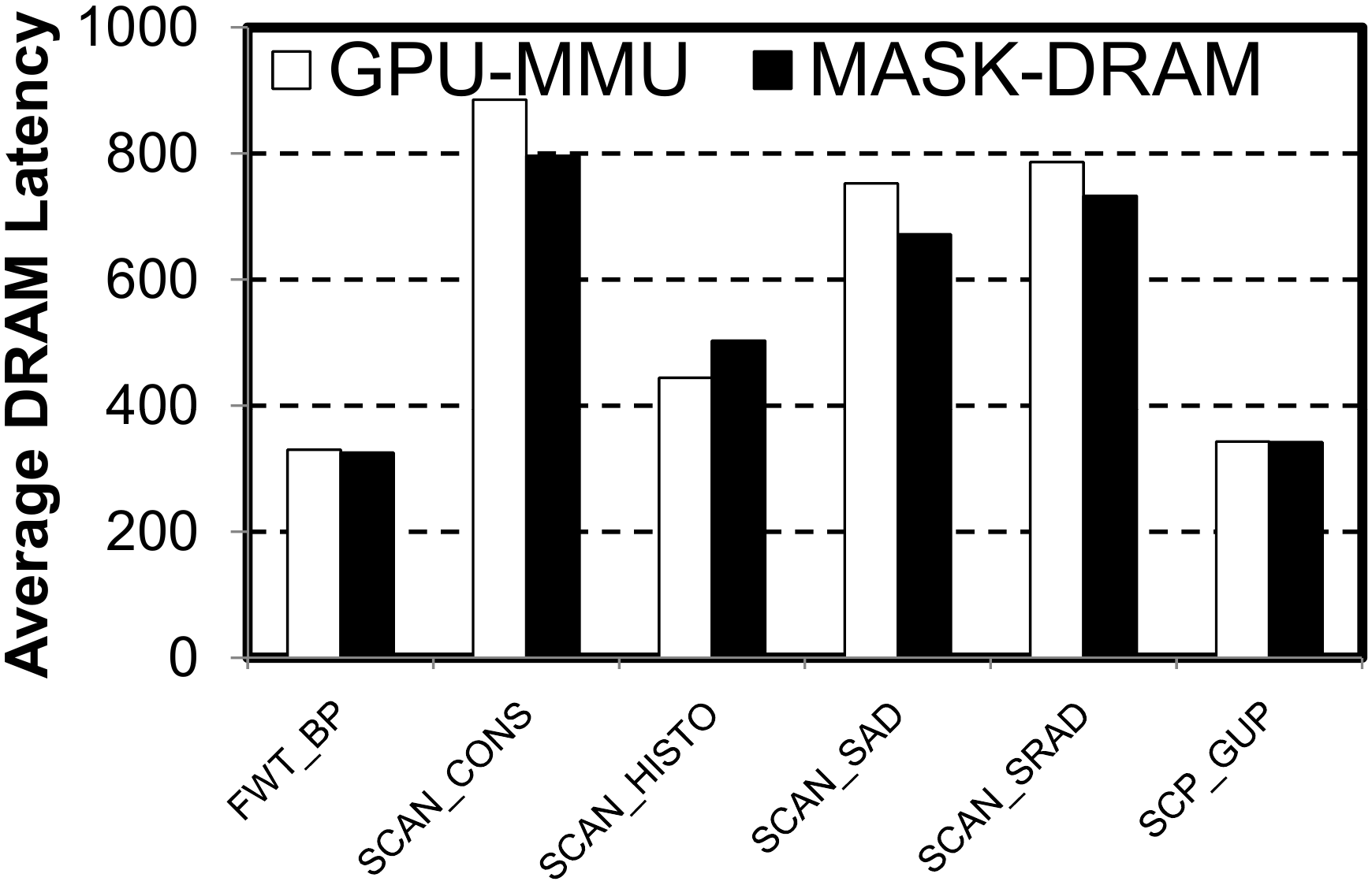}%
	\label{fig:multi-dram-lat}
}}%
}
\caption{DRAM bandwidth utilization and latency.\cjr{We could save space by 
		combining legends for all these graphs.}}

\label{fig:multi-dram}
\end{figure}

\subsection{Scalability and Performance on Other Architectures}
\label{sec:750-eval}

Figure~\ref{fig:eval-scaling} shows the performance of GPU-MMU and \titleShort,
normalized to the ideal performance with no translation overhead, as we vary the
number of applications executing concurrently on the GPU. We observe that
as the application count increases, the performance of both GPU-MMU and \titleShort
are further from the ideal baseline, due to contention for shared
resources (e.g., shared TLB, shared data cache). However, \titleShort provides
increasingly better performance compared to GPU-MMU (35.5\% for one
application, 45.2\% for two concurrent applications, and 47.3\% for three
concurrent applications). We conclude that \titleShort provides better
scalability with application count over the state-of-the-art designs.

\begin{figure}[h]
\centering
\resizebox{\columnwidth}{!}{%
\subfloat[Scalability analysis]{{
	\includegraphics[width=0.4\columnwidth,valign=t]{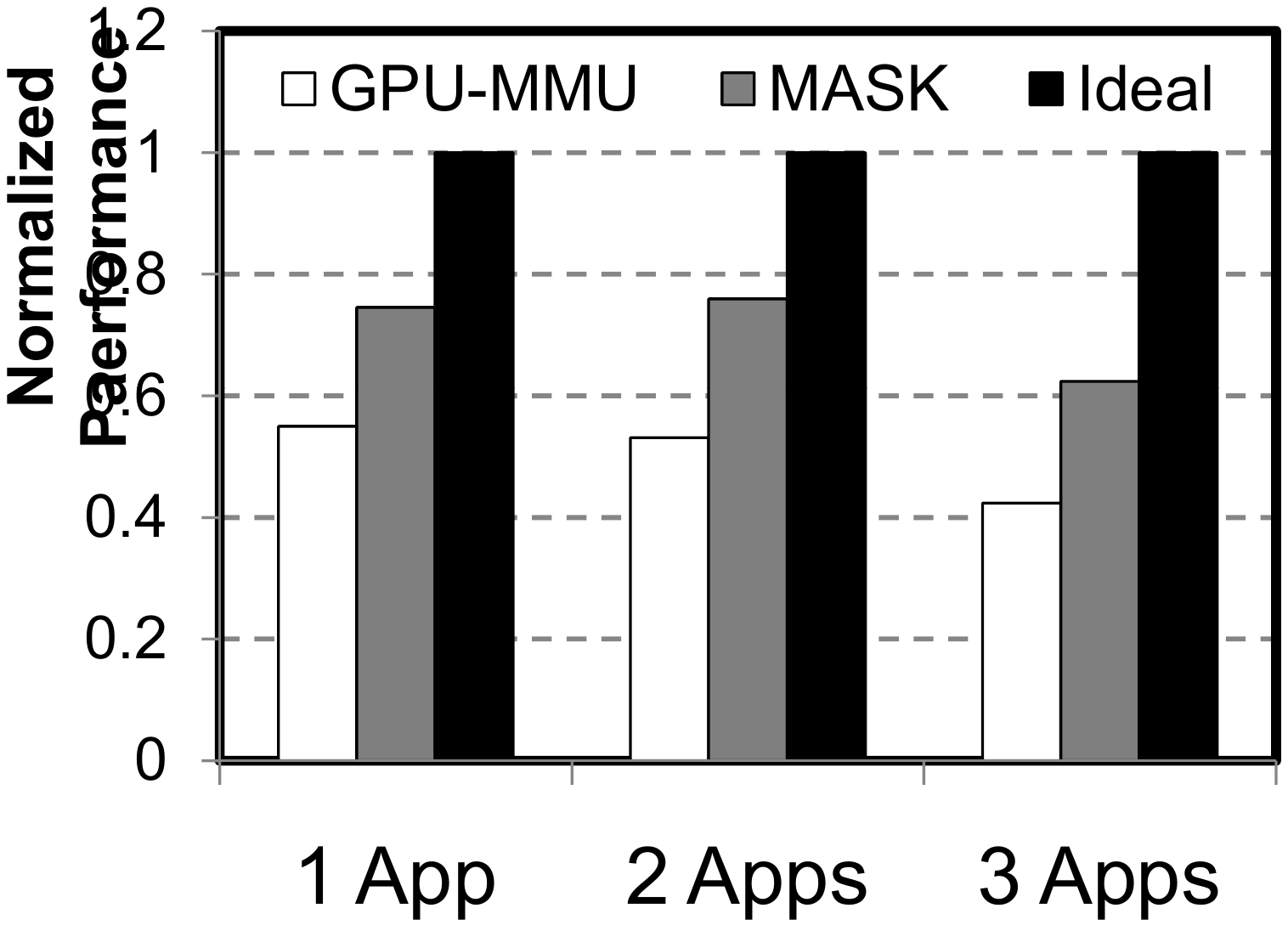}%
	\vphantom{\includegraphics[width=0.4\columnwidth,valign=t]{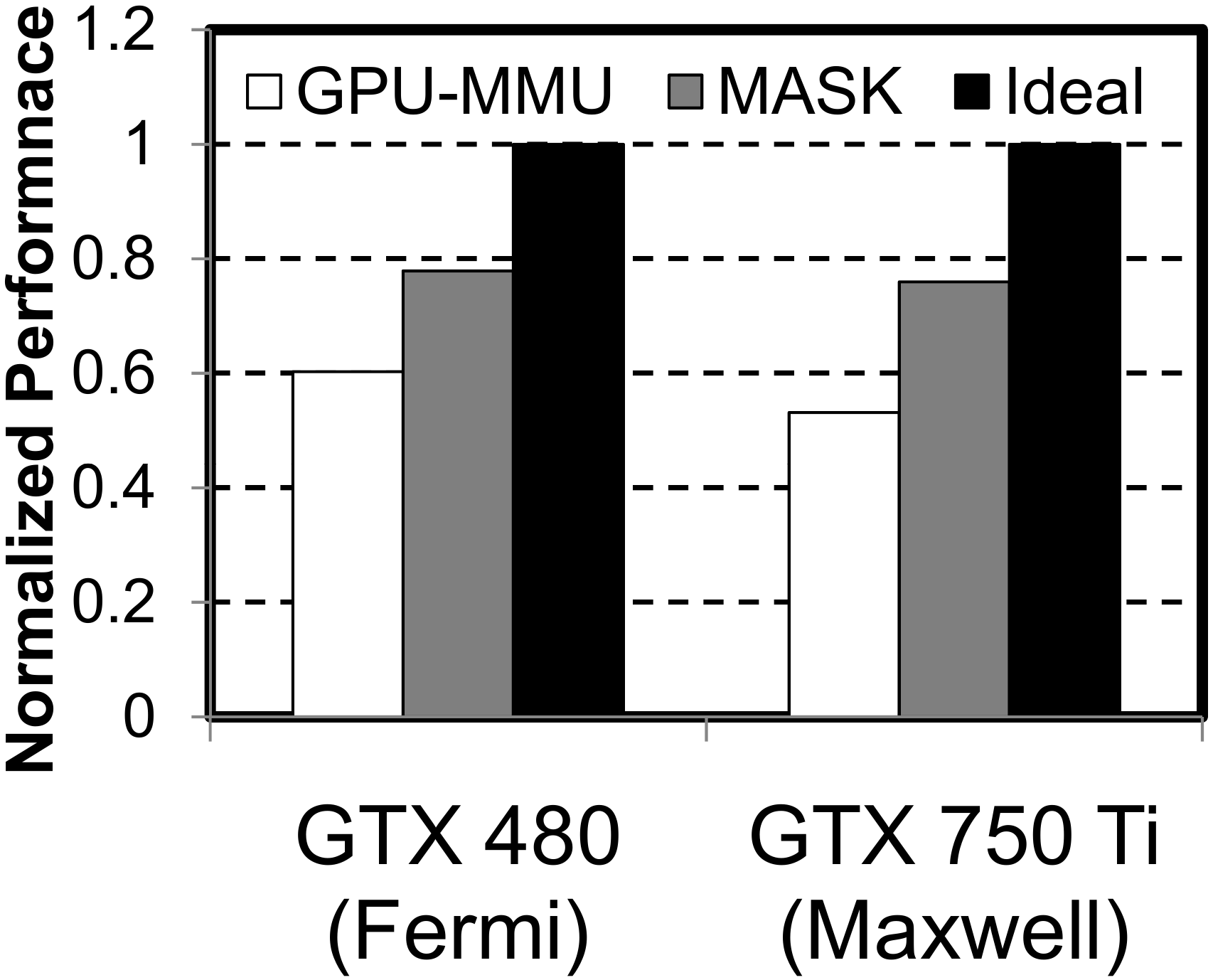}}%
	\label{fig:eval-scaling}
}}%

\subfloat[Performance on Fermi]{{
\includegraphics[width=0.4\columnwidth,valign=t]{./perf-arch.pdf}%
\label{fig:eval-maxwell}
}}%
}
\caption{Scalability and portability studies for \titleShort.}
\label{fig:mask-scaling-and-arch}
\end{figure}

The analyses and designs of \titleShort{} are architecture independent and
should be applicable to any SIMD machine. To demonstrate this, we evaluate
\titleShort on the GTX 480, which uses the Fermi architecture~\cite{fermi}.
Figure~\ref{fig:eval-maxwell} shows the performance of GPU-MMU and
\titleShort{}, normalized to the ideal performance with no translation
overhead, for the GTX 480 and the GTX 750 Ti. We make three observations.
First, address translation incurs significant performance overhead in both
architectures for the baseline GPU-MMU design.  Second, \titleShort{} provides
a 29.1\% performance improvement over the GPU-MMU design in the Fermi
architecture.  Third, compared to the ideal performance, \titleShort{} performs
only 22\% worse in the Fermi architecture. On top of the data shown in
Figure~\ref{fig:eval-maxwell}, we find that \titleShort{} reduces unfairness
by 26.4\% and increases the TLB hit rates by 64.7\% on average the in Fermi
architecture. We conclude that \titleShort{} delivers significant benefits
regardless of GPU architecture.

Aside from this, Table~\ref{fig:mask-pwcache-sweep} provides an
evaluation of \titleShort{} on the integrated GPU configuration used in
previous work~\cite{powers-hpca14}.  This integrated GPU has fewer number of
GPU cores, slower L2 cache, slower and less bandwidth main memory.

\begin{table}[h!]
\begin{scriptsize}
  \centering
    \begin{tabular}{|c|c|c|}
\hline
\textbf{Relative Performance} & \textbf{Maxwell} & \textbf{Integrated GPU~\cite{powers-hpca14}}  \\ \hline
\hline
\textbf{Shared TLB} & 52.4\% & 38.2\%                                                         \\ \hline
\textbf{\titleShort + Shared TLB} & 76.3\% & 64.5\%                                            \\ \hline
\textbf{Translation Cache} & 46.0\% & 52.1\%                                                  \\ \hline
\textbf{\titleShort + Translation Cache} &  72.6\% & 72.5\%                                      \\ \hline
    \end{tabular}%
  \caption{\footnotesize{Relative performance vs. the ideal baseline.}}
  \label{fig:mask-pwcache-sweep}
\end{scriptsize}%
\end{table}%

From Table~\ref{fig:mask-pwcache-sweep}, we found that 1) \titleShort{} is
effective in reducing the latency of address translation and able to improve
the performance of both the shared L2 TLB and translation cache designs on both
off-chip Maxwell GPU and integrated GPU configurations, 2) contention
at the shared L2 TLB becomes significantly more severe and causes a significant 
performance drop in the integrated GPU setup.


\subsection{Sensitivity Studies}
\label{sec:eval-sense}

\para{Sensitivity to L1 and L2 TLB Sizes.}
We evaluated the performance of
\titleShort{} for a range of L1 and L2 TLB sizes. We find that for both the L1 and L2 TLB, 
\titleShort{} performs closer to the baseline as the number of TLB entries increases,
as the contention at the L1 and L2 TLB decreases.


\para{Sensitivity to Memory Policies.} We study the sensitivity
of \titleShort{} to (1)~main memory row policy, and (2)~memory scheduling
policies. We find that for both the GPU-MMU baseline and \titleShort{}, the
workload performance for an open-row policy is similar (within 0.8\%) when we
instead employ a closed row policy, which is used in various CPU
processors~\cite{ivybridge,intel-sandybridge,skylake}. Aside from the FR-FCFS
scheduler~\cite{fr-fcfs,frfcfs-patent}, we applied \titleShort on other
state-of-the-art GPU memory scheduler~\cite{mafia} and found that \titleShort{}
with this scheduler performs 44.2\% over the GPU-MMU baseline. We conclude that
\titleShort is effective across different memory policies. 

\para{Sensitivity to Different Page Size.} We evaluate the performance of
\titleShort with large page assuming ideal page fault latency. We found that applying
\titleShort{} allows the GPU to perform within 1.8\% of the ideal baseline.

\subsection{Hardware Overheads}
\label{sec:overhead}

To support memory protection, each L2 TLB cache line adds an address space identifier (ASID).
We model 8-bit ASIDs added to TLB entries, which translates to 7\% of the L2 TLB size. 

\tlbtokenname, uses two 16-bit counters at each shader core.
We augment the shared cache with
32-entry fully-associative content addressable memory (CAM) for the bypass
cache, and 30 15-bit token counts with 30 1-bit token direction entries to distribute
tokens over up to 30 concurrent applications. In total, we add 436 bytes
(4 bytes per core on the L1 TLB, and 316 bytes in the
shared L2 TLB), which represents 0.5\% growth of the L1 TLB and
3.8\% of the L2 TLB.

\cachebypass uses ten 8-byte counters per core to track
cache hits and cache accesses per level, (including for the data cache).
The resulting 80 bytes are less than 0.1\% 
of the shared L2 cache. Each cache and memory request requires an
additional 3 bits
specifying the page walk level, as discussed in
Section~\ref{sec:data-bypassing}.

\dramsched adds a 16-entry FIFO queue in each memory channel for TLB-related
requests, and a 64-entry memory request buffer per memory channel for the
\silverQ, while reducing the size of the \normalQ by 64 entries down to 192
entries. This adds an extra 6\% of storage overhead to the DRAM request queue per
memory controller.

\para{Area and Power Consumption.} We compare the area and power
consumption of \titleShort{} using
CACTI~\cite{cacti}. We compare the area and power of the L1 TLB, L2 TLB, the shared data cache and the page walk cache. We
find that \titleShort{} introduces a negligible overhead, consuming less than
0.1\% additional area and 0.01\% additional power than both shared L2 TLB and page 
walk cache baselines.

%
%
%
%
%

\section{Related Work}

\subsection{Partitioning for GPU Concurrency}

\para{Concurrent Kernels and GPU Multiprogramming.} 
The opportunity to improve utilization 
with concurrency is well-recognized, but previous 
proposals~\cite{asplos-sree,wang-hpca16,li2014symbiotic,pagoda-ppopp17}
do not support memory protection.
Adriaens et al.~\cite{gpu-multitasking} 
observe the need for spatial sharing across protection domains,
but do not propose or evaluate a design.
NVIDIA GRID~\cite{grid} and
AMD Firepro~\cite{firepro} support static
partitioning of hardware to allow kernels from different VMs
to run concurrently, but the partitions are determined at startup,
which causes fragmentation and under-utilization (see Section~\ref{sec:eval-multi}).
\titleShort{}'s goal is \emph{flexible, dynamic} partitioning. 

NVIDIA's Multi Process Service (MPS)~\cite{mps} allows multiple processes to 
launch kernels on the GPU, but the service provides no memory protection or error
containment. 
Xu et al.~\cite{warp-slicer} propose Warped-Slicer,
which is a mechanism for multiple applications to spatially share a GPU core.
Similar to MPS, Warped-Slicer provides no memory protection, and is not suitable
for supporting multi-application in a multi-tenant cloud setting.

\para{Preemption and Context Switching.}
Preemptive context switch 
is an active research area~\cite{isca-2014-preemptive, gebhart,wang-hpca16}, and 
architectural support~\cite{lindholm,pascal} will likely improve in future GPUs.
Preemption is complementary to spatial multiplexing,
and we leave techniques to combine them for future work.

\para{GPU Virtualization.} 
Most current hypervisor-based full virtualization techniques for GPGPUs~\cite{gdev,gpuvm,gVirt} 
must support a virtual device abstraction without the dedicated hardware
support for the Virtual Desktop Infrastructure (VDI) found in GRID~\cite{grid} and FirePro~\cite{firepro}. Key components missing from these
proposals include support for the dynamic partitioning of
hardware resources, and efficient techniques for handling over-subscription. Performance
overheads incurred by these designs argue strongly for hardware assistance,
as we propose. By contrast, API-remoting solutions such as vmCUDA~\cite{vmCUDA} and
rCUDA~\cite{rcuda} provide near-native performance, but require modifications
to the guest software and sacrifice both isolation and compatibility.

\para{Demand Paging in GPUs.}
Demand paging is an important primitive for memory virtualization that 
is challenging for GPUs~\cite{abhishek-ispass16}. 
Recent works on CC-NUMA~\cite{cc-numa-gpu-hpca15}, AMD's hUMA~\cite{huma}, and
NVIDIA's PASCAL architecture~\cite{tianhao-hpca16,pascal}
support for demand paging in GPUs. These techniques can be used in conjunction
with \titleShort{}.


\subsection{TLB Design}

\para{GPU TLB Designs.} Previous works have explored TLB designs in
heterogeneous systems with
GPUs~\cite{cong-hpca17,powers-hpca14,pichai-asplos14,abhishek-ispass16}, and the
adaptation of x86-like TLBs in a heterogeneous CPU-GPU
setting~\cite{powers-hpca14}.  Key elements in these designs include probing
the TLB after L1 coalescing to reduce the amount of parallel TLB requests,
shared concurrent page table walks, and translation caches to reduce main
memory accesses. \titleShort{} owes much to these designs, but we show
empirically that contention patterns at the shared L2 layer require additional
support beyond these designs to accommodate contention from multiple address spaces. Cong et al. propose a TLB
design similar to our baseline GPU-MMU design~\cite{cong-hpca17}.  However,
this design utilizes the host (CPU) MMU to perform page walks, which is
inapplicable in the context of multi-application GPUs. Pichai et
al.~\cite{pichai-asplos14} explore a TLB design for heterogeneous CPU-GPU systems,
and add TLB awareness to the existing CCWS GPU warp scheduler~\cite{ccws}.
Warp scheduling is orthogonal to our work, and 
can be combined to further improve performance.

Vesely et al.\ analyze support for virtual memory in heterogeneous
systems~\cite{abhishek-ispass16}, finding that the cost of address translation
in GPUs is an order of magnitude higher than in CPUs, and that high latency
address translations limit the GPU's latency hiding capability and hurts
performance (an observation in line with our own findings in
Section~\ref{sec:tlb-bottleneck}). We show additionally that thrashing due to
interference further slows down applications sharing the GPU. \titleShort is capable
not only of reducing interference between multiple applications
(Section~\ref{sec:eval-multi}), but of reducing the TLB miss rate in
single-application scenarios as well.

Instead of relying on hardware modifications, Lee et al.\ propose VAST, a
software-managed virtual memory space for GPUs~\cite{vast}. 
Data-parallel applications typically have a larger working set size compared to
the size of GPU memory, preventing these applications from utilizing the GPUs.
To address this, VAST creates the illusion of a large virtual memory (without
concerns about the physical memory size), by providing an automatic memory
management system that partitions GPU programs into chunks that fit the physical
memory space. Even though recent GPUs now support demand paging~\cite{pascal},
the observation regarding the large working set size of GPGPU programs
motivates the need for better virtual memory support, which is what \titleShort provide.

\para{TLB Designs in CPU Systems.} 
Cox and Bhattacharjee propose an efficient TLB deign that allows entries
corresponding to multiple page sizes to share the same TLB structure,
simplifying the design of TLBs~\cite{cox-asplos16}. While this design can be applied to GPUs, it is
solving a different problem: area and energy efficiency. Thus, this proposal
is orthogonal to \titleShort{}.  Bhattacharjee et al.\ examine shared last-level
TLB designs~\cite{inter-core-tlb} and page walk cache
designs~\cite{large-reach}, proposing a mechanism that can accelerate
multithreaded applications by sharing translations between cores. However,
these proposals are likely to be less effective for multiple concurrent GPGPU
applications, because translations are not shared between virtual address
spaces.  Barr et al.\ propose SpecTLB~\cite{spectlb}, which speculatively
predicts address translations to avoid the TLB miss latency.  Speculatively
predicting address translation can be complicated and costly in GPUs, because
there can be multiple concurrent TLB misses to many different TLB entries in
the GPU.

Direct segments~\cite{direct-segment} and redundant memory mappings~\cite{rmm}
reduce address translation overheads by mapping large contiguous virtual memory regions
to a contiguous physical region. These techniques increase
 the reach of each TLB entry, and are
complementary to those in \titleShort{}.


\subsection{Techniques to Reduce Interference}


\para{GPU-Specific Resource Management.}
Jog et al.\ propose MAFIA, a main memory management
scheme that improves performance of concurrently-running GPGPU
applications~\cite{mafia}. The design of MAFIA assumes that parallel applications
operate under the same virtual address space, and does not model address
translation overheads or accommodate {\em safe}, concurrent execution of kernels
from different protection domains. In contrast, we model and study 
the impact of address translation and memory protection. 
Lee et al.\ propose TAP~\cite{lee2012tap}, a TLP-aware cache
management mechanism that modifies the CPU cache partitioning
policy~\cite{ucp-yale} and cache insertion policy~\cite{rrip} to lower GPGPU
applications' interference to CPU applications at the shared cache. However,
TAP does not consider address translation and interference
between different GPGPU applications.

Several memory scheduler designs target systems with
GPUs~\cite{complexity,chatterjee-sc14,sms,jeong-cpu-gpu,usui-squash,usui-dash}. Unlike
\titleShort{}, these designs focus on a single GPGPU application, and are not
aware of page walk traffic. They focus on reducing the complexity of the memory
scheduler for a single application by reducing
inter-warp interference~\cite{complexity, chatterjee-sc14}, or by
providing resource management for heterogeneous CPU-GPU
applications~\cite{sms,jeong-cpu-gpu,usui-squash,usui-dash}. While some of these works propose
mechanisms that reduce interference~\cite{sms,jeong-cpu-gpu,usui-squash,usui-dash}, they 
differ from \titleShort because 1)~they consider interference from
applications with wildly different characteristics (CPU applications vs. GPU
applications), and 2)~they do not consider interference between page-walk-related
and normal memory traffic.


\para{Cache Bypassing Policies in GPUs.} Techniques to reduce
contention on shared GPU
caches~\cite{medic,donglihpca15,li-ics15,xie-hpca15,chen-micro47,chen-mes14} employ
memory-divergence-based bypassing~\cite{medic}, reuse-based cache
bypassing~\cite{donglihpca15,li-ics15,xie-hpca15,chen-micro47,chen-mes14}, and software-based
cache bypassing~\cite{xie-iccad13}, and sometimes combine these works
with throttling~\cite{chen-micro47,chen-mes14,mrpb} to reduce contention. These
works do not differentiate page walk traffic from normal traffic, and focus on a single 
application. 


\para{Cache and TLB Insertion Policies.} Cache
insertion policies that account for cache thrashing~\cite{dip,rrip,bip} or
future reuse~\cite{eaf-vivek} work well for CPU
applications, but other previous works have shown these policies to be
ineffective for GPU applications~\cite{lee2012tap,medic}. 
This observation holds for the shared TLB in the multi-address space scenario.

\section{Conclusion}

Efficiently deploying GPUs in a large-scale computing environment needs spatial multiplexing support.
However, the existing
address translation support stresses a GPU's fundamental latency 
hiding techniques, and interference from multiple address spaces can further harm performance.
To alleviate these problems, we propose \titleShort{},
a new memory hierarchy designed for multi-address-space concurrency.
\titleShort{} consists of three major components that lower inter-application 
interference during address translation and improve L2 cache utilization
for translation requests.
\titleShort{} successfully alleviates the address translation overhead,
improving performance by 45.2\% 
over the state-of-the-art.


\bibliographystyle{IEEEtranS}
\bibliography{references}

\begin{thebibliography}{10}
\providecommand{\url}[1]{#1}
\csname url@samestyle\endcsname
\providecommand{\newblock}{\relax}
\providecommand{\bibinfo}[2]{#2}
\providecommand{\BIBentrySTDinterwordspacing}{\spaceskip=0pt\relax}
\providecommand{\BIBentryALTinterwordstretchfactor}{4}
\providecommand{\BIBentryALTinterwordspacing}{\spaceskip=\fontdimen2\font plus
\BIBentryALTinterwordstretchfactor\fontdimen3\font minus
  \fontdimen4\font\relax}
\providecommand{\BIBforeignlanguage}[2]{{%
\expandafter\ifx\csname l@#1\endcsname\relax
\typeout{** WARNING: IEEEtranS.bst: No hyphenation pattern has been}%
\typeout{** loaded for the language `#1'. Using the pattern for}%
\typeout{** the default language instead.}%
\else
\language=\csname l@#1\endcsname
\fi
#2}}
\providecommand{\BIBdecl}{\relax}
\BIBdecl

\bibitem{tensorflow}
\BIBentryALTinterwordspacing
M.~Abadi, A.~Agarwal, P.~Barham, E.~Brevdo, Z.~Chen, C.~Citro, G.~Corrado,
  A.~Davis, J.~Dean, M.~Devin, S.~Ghemawat, I.~Goodfellow, A.~Harp, G.~Irving,
  M.~Isard, Y.~Jia, R.~Jozefowicz, L.~Kaiser, M.~Kudlur, J.~Levenberg,
  D.~Mané, R.~Monga, S.~Moore, D.~Murray, C.~Olah, M.~Schuster, J.~Shlens,
  B.~Steiner, I.~Sutskever, K.~Talwar, P.~Tucker, V.~Vanhoucke, V.~Vasudevan,
  F.~Viégas, O.~Vinyals, P.~Warden, M.~Wattenberg, M.~Wicke, Y.~Yu, and
  X.~Zheng, ``{TensorFlow: Large-Scale Machine Learning on Heterogeneous
  Distributed Systems},'' 2015. [Online]. Available:
  \url{http://download.tensorflow.org/paper/whitepaper2015.pdf}
\BIBentrySTDinterwordspacing

\bibitem{gpu-multitasking}
J.~Adriaens, K.~Compton, N.~S. Kim, and M.~Schulte, ``{The Case for GPGPU
  Spatial Multitasking},'' in \emph{HPCA}, 2012.

\bibitem{firepro}
{Advanced Micro Devices}, ``{OpenCL: The Future of Accelerated Application
  Performance Is Now},''
  \url{https://www.amd.com/Documents/FirePro_OpenCL_Whitepaper.pdf}.

\bibitem{npt}
{Advanced Micro Devices}, \emph{{AMD-V Nested Paging}}, 2010,
  \url{http://developer.amd.com/wordpress/media/2012/10/NPT-WP-1%201-final-TM.pdf}.

\bibitem{huma}
{Advanced Micro Devices}, ``{Heterogeneous System Architecture: A Technical
  Review},''
  \url{http://amd-dev.wpengine.netdna-cdn.com/wordpress/media/2012/10/hsa10.pdf},
  2012.

\bibitem{amd-hsa}
\BIBentryALTinterwordspacing
{Advanced Micro Devices}. ({2013}) {What is Heterogeneous System Architecture
  (HSA)?} [Online]. Available:
  \url{{http://developer.amd.com/resources/heterogeneous-computing/what-is-heterogeneous-system-architecture-hsa/}}
\BIBentrySTDinterwordspacing

\bibitem{amd-io-virt}
\BIBentryALTinterwordspacing
{Advanced Micro Devices}, ``{AMD I/O Virtualization Technology (IOMMU)
  Specification},'' 2016. [Online]. Available:
  \url{{http://support.amd.com/TechDocs/48882_IOMMU.pdf}}
\BIBentrySTDinterwordspacing

\bibitem{cc-numa-gpu-hpca15}
N.~Agarwal, D.~Nellans, M.~O'Connor, S.~W. Keckler, and T.~F. Wenisch,
  ``{Unlocking Bandwidth for GPUs in CC-NUMA Systems},'' in \emph{HPCA}, 2015.

\bibitem{amit10wiosca}
N.~Amit, M.~Ben-Yehuda, and B.-A. Yassour, ``{IOMMU: Strategies for Mitigating
  the IOTLB Bottleneck},'' in \emph{ISCA}, 2012.

\bibitem{arunkumar-isca17}
A.~Arunkumar, E.~Bolotin, B.~Cho, U.~Milic, E.~Ebrahimi, O.~Villa, A.~Jaleel,
  and C.-J. Wu, ``{MCM-GPU: Multi-Chip-Module GPUs for Continued Performance
  Scalability},'' in \emph{ISCA}, 2017.

\bibitem{sms}
R.~Ausavarungnirun, K.~Chang, L.~Subramanian, G.~Loh, and O.~Mutlu, ``{Staged
  Memory Scheduling: Achieving High Performance and Scalability in
  Heterogeneous Systems},'' in \emph{ISCA}, 2012.

\bibitem{medic}
R.~Ausavarungnirun, S.~Ghose, O.~Kayıran, G.~H. Loh, C.~R. Das, M.~T.
  Kandemir, and O.~Mutlu, ``{Exploiting Inter-Warp Heterogeneity to Improve
  GPGPU Performance},'' in \emph{PACT}, 2015.

\bibitem{gpgpu-sim}
A.~Bakhoda, G.~Yuan, W.~Fung, H.~Wong, and T.~Aamodt, ``{Analyzing CUDA
  Workloads Using a Detailed GPU Simulator},'' in \emph{ISPASS}, 2009.

\bibitem{spectlb}
T.~W. Barr, A.~L. Cox, and S.~Rixner, ``{SpecTLB: A Mechanism for Speculative
  Address Translation},'' in \emph{ISCA}, 2011.

\bibitem{direct-segment}
A.~Basu, J.~Gandhi, J.~Chang, M.~D. Hill, and M.~M. Swift, ``{Efficient Virtual
  Memory for Big Memory Servers},'' in \emph{ISCA}, 2013.

\bibitem{large-reach}
A.~Bhattacharjee, ``{Large-reach Memory Management Unit Caches},'' in
  \emph{MICRO}, 2013.

\bibitem{inter-core-tlb}
A.~Bhattacharjee and M.~Martonosi, ``{Inter-core Cooperative TLB for Chip
  Multiprocessors},'' in \emph{ASPLOS}, 2010.

\bibitem{tlb-consistency}
D.~L. Black, R.~F. Rashid, D.~B. Golub, and C.~R. Hill, ``{Translation
  Lookaside Buffer Consistency: A Software Approach},'' in \emph{ASPLOS}, 1989.

\bibitem{kaveri}
D.~Bouvier and B.~Sander, ``{Applying AMD's "Kaveri" APU for Heterogeneous
  Computing},'' in \emph{HOTCHIP}, 2014.

\bibitem{lonestar}
M.~Burtscher, R.~Nasre, and K.~Pingali, ``{A Quantitative Study of Irregular
  Programs on {GPUs}},'' in \emph{IISWC}, 2012.

\bibitem{chatterjee-sc14}
N.~Chatterjee, M.~O'Connor, G.~H. Loh, N.~Jayasena, and R.~Balasubramonian,
  ``{Managing DRAM Latency Divergence in Irregular GPGPU Applications},'' in
  \emph{SC}, 2014.

\bibitem{rodinia}
S.~Che, M.~Boyer, J.~Meng, D.~Tarjan, J.~Sheaffer, S.-H. Lee, and K.~Skadron,
  ``{Rodinia: A Benchmark Suite for Heterogeneous Computing},'' in
  \emph{IISWC}, 2009.

\bibitem{chen-micro47}
X.~Chen, L.-W. Chang, C.~I. Rodrigues, J.~Lv, Z.~Wang, and W.~W. Hwu,
  ``{Adaptive Cache Management for Energy-Efficient {GPU} Computing},'' in
  \emph{MICRO}, 2014.

\bibitem{chen-mes14}
X.~Chen, S.~Wu, L.-W. Chang, W.-S. Huang, C.~Pearson, Z.~Wang, and W.~W. Hwu,
  ``{Adaptive Cache Bypass and Insertion for Many-Core Accelerators},'' in
  \emph{MES}, 2014.

\bibitem{cong-hpca17}
J.~Cong, Z.~Fang, Y.~Hao, and G.~Reinmana, ``{Supporting Address Translation
  for Accelerator-Centric Architectures},'' in \emph{HPCA}, 2017.

\bibitem{cox-asplos16}
G.~Cox and A.~Bhattacharjee, ``{Efficient Address Translation with Multiple
  Page Sizes},'' in \emph{ASPLOS}, 2016.

\bibitem{shoc}
A.~Danalis, G.~Marin, C.~McCurdy, J.~S. Meredith, P.~C. Roth, K.~Spafford,
  V.~Tipparaju, and J.~S. Vetter, ``{The Scalable Heterogeneous Computing
  (SHOC) benchmark suite},'' in \emph{GPGPU}, 2010.

\bibitem{rcuda}
J.~Duato, A.~Pena, F.~Silla, R.~Mayo, and E.~Quintana-Orti, ``{rCUDA: Reducing
  the Number of GPU-based Accelerators in High Performance Clusters},'' in
  \emph{HPCS}, 2010.

\bibitem{ebrahimi-micro09}
E.~Ebrahimi, O.~Mutlu, C.~J. Lee, and Y.~N. Patt, ``{Coordinated Control of
  Multiple Prefetchers in Multi-core Systems},'' in \emph{MICRO}, 2009.

\bibitem{harmonic_speedup}
S.~Eyerman and L.~Eeckhout, ``{System-Level Performance Metrics for
  Multiprogram Workloads},'' \emph{IEEE Micro}, vol.~28, no.~3, 2008.

\bibitem{ws-metric2}
S.~Eyerman and L.~Eeckhout, ``{Restating the Case for Weighted-IPC Metrics to
  Evaluate Multiprogram Workload Performance},'' \emph{IEEE CAL}, 2014.

\bibitem{flynn}
M.~Flynn, ``{Very High-Speed Computing Systems},'' \emph{Proc.\ of the IEEE},
  vol.~54, no.~2, 1966.

\bibitem{gebhart}
M.~Gebhart, D.~R. Johnson, D.~Tarjan, S.~W. Keckler, W.~J. Dally, E.~Lindholm,
  and K.~Skadron, ``{Energy-Efficient Mechanisms for Managing Thread Context in
  Throughput Processors},'' in \emph{ISCA}, 2011.

\bibitem{mars}
B.~He, W.~Fang, Q.~Luo, N.~K. Govindaraju, and T.~Wang, ``{Mars: A MapReduce
  Framework on Graphics Processors},'' in \emph{PACT}, 2008.

\bibitem{grid}
A.~Herrera, ``{NVIDIA GRID: Graphics Accelerated VDI with the Visual
  Performance of a Workstation},'' May 2014.

\bibitem{intel-sandybridge}
{Intel Corporation}, ``{Intel(R) Microarchitecture Codename Sandy Bridge},''
  http://www.intel.com/technology/architecture-silicon/2ndgen/.

\bibitem{ivybridge}
\BIBentryALTinterwordspacing
{Intel Corporation}. ({2012}) {Products (Formerly Ivy Bridge)}. [Online].
  Available: \url{{http://ark.intel.com/products/codename/29902/}}
\BIBentrySTDinterwordspacing

\bibitem{ept}
{Intel Corporation}, ``Intel 64 and ia-32 architectures software developer’s
  manual,'' 2016,
  \url{https://www-ssl.intel.com/content/dam/www/public/us/en/documents/manuals/64-ia-32-architectures-software-developer-manual-325462.pdf}.

\bibitem{intel-io-virt}
\BIBentryALTinterwordspacing
{Intel Corporation}, ``Intel virtualization technology for directed i/o,''
  2016. [Online]. Available:
  \url{{http://www.intel.com/content/dam/www/public/us/en/documents/product-specifications/vt-directed-io-spec.pdf}}
\BIBentrySTDinterwordspacing

\bibitem{skylake}
{Intel Corporation}, ``6th generation intel® core™ processor family
  datasheet, vol. 1,'' 2017,
  \url{http://www.intel.com/content/dam/www/public/us/en/documents/datasheets/desktop-6th-gen-core-family-datasheet-vol-1.pdf}.

\bibitem{dip}
A.~Jaleel, W.~Hasenplaugh, M.~Qureshi, J.~Sebot, S.~Steely, Jr., and J.~Emer,
  ``{Adaptive Insertion Policies for Managing Shared Caches},'' in \emph{PACT},
  2008.

\bibitem{rrip}
A.~Jaleel, K.~B. Theobald, S.~C. Steely, Jr., and J.~Emer, ``{High Performance
  Cache Replacement Using Re-reference Interval Prediction ({RRIP})},'' in
  \emph{ISCA}, 2010.

\bibitem{jeong-cpu-gpu}
M.~K. Jeong, M.~Erez, C.~Sudanthi, and N.~Paver, ``{A QoS-aware memory
  controller for dynamically balancing GPU and CPU bandwidth use in an
  MPSoC},'' in \emph{DAC}, 2012.

\bibitem{mrpb}
W.~Jia, K.~A. Shaw, and M.~Martonosi, ``{{MRPB}: Memory Request Prioritization
  for Massively Parallel Processors},'' in \emph{HPCA}, 2014.

\bibitem{mafia}
A.~Jog, O.~Kayiran, T.~Kesten, A.~Pattnaik, E.~Bolotin, N.~Chatterjee, S.~W.
  Keckler, M.~T. Kandemir, and C.~R. Das, ``{Anatomy of GPU Memory System for
  Multi-Application Execution},'' in \emph{MEMSYS}, 2015.

\bibitem{rmm}
V.~Karakostas, J.~Gandhi, F.~Ayar, A.~Cristal, M.~D. Hill, K.~S. McKinley,
  M.~Nemirovsky, M.~M. Swift, and O.~\"{U}nsal, ``{Redundant Memory Mappings
  for Fast Access to Large Memories},'' in \emph{ISCA}, 2015.

\bibitem{lulesh}
I.~Karlin, A.~Bhatele, J.~Keasler, B.~Chamberlain, J.~Cohen, Z.~DeVito,
  R.~Haque, D.~Laney, E.~Luke, F.~Wang, D.~Richards, M.~Schulz, and C.~Still,
  ``{Exploring Traditional and Emerging Parallel Programming Models using a
  Proxy Application},'' in \emph{IPDPS}, 2013.

\bibitem{lulesh-origin}
I.~Karlin, J.~Keasler, and R.~Neely, ``{Lulesh 2.0 Updates and Changes},''
  2013.

\bibitem{gdev}
S.~Kato, M.~McThrow, C.~Maltzahn, and S.~Brandt, ``{Gdev: First-Class GPU
  Resource Management in the Operating System},'' in \emph{USENIX ATC}, 2012.

\bibitem{nmnl-pact13}
O.~Kay{\i}ran, A.~Jog, M.~T. Kandemir, and C.~R. Das, ``{Neither More Nor Less:
  Optimizing Thread-Level Parallelism for GPGPUs},'' in \emph{PACT}, 2013.

\bibitem{cpugpu-micro}
O.~Kay{\i}ran, N.~C. Nachiappan, A.~Jog, R.~Ausavarungnirun, M.~T. Kandemir,
  G.~H. Loh, O.~Mutlu, and C.~R. Das, ``{Managing GPU Concurrency in
  Heterogeneous Architectures},'' in \emph{MICRO}, 2014.

\bibitem{lee2012tap}
J.~Lee and H.~Kim, ``Tap: A tlp-aware cache management policy for a cpu-gpu
  heterogeneous architecture,'' in \emph{High Performance Computer Architecture
  (HPCA), 2012 IEEE 18th International Symposium on}.\hskip 1em plus 0.5em
  minus 0.4em\relax IEEE, 2012, pp. 1--12.

\bibitem{vast}
J.~Lee, M.~Samadi, and S.~Mahlke, ``{VAST: The Illusion of a Large Memory Space
  for GPUs},'' in \emph{PACT}, 2014.

\bibitem{li-ics15}
C.~Li, S.~L. Song, H.~Dai, A.~Sidelnik, S.~K.~S. Hari, and H.~Zhou,
  ``{Locality-Driven Dynamic {GPU} Cache Bypassing},'' in \emph{ICS}, 2015.

\bibitem{donglihpca15}
D.~Li, M.~Rhu, D.~Johnson, M.~O'Connor, M.~Erez, D.~Burger, D.~Fussell, and
  S.~Redder, ``{Priority-Based Cache Allocation in Throughput Processors},'' in
  \emph{HPCA}, 2015.

\bibitem{li2014symbiotic}
T.~Li, V.~K. Narayana, and T.~El-Ghazawi, ``{Symbiotic Scheduling of Concurrent
  GPU Kernels for Performance and Energy Optimizations},'' in \emph{CF}, 2014.

\bibitem{lindholm}
E.~Lindholm, J.~Nickolls, S.~Oberman, and J.~Montrym, ``{NVIDIA Tesla: A
  Unified Graphics and Computing Architecture},'' \emph{IEEE Micro}, vol.~28,
  no.~2, 2008.

\bibitem{igpu}
J.~Menon, M.~de~Kruijf, and K.~Sankaralingam, ``{iGPU: Exception Support and
  Speculative Execution on GPUs},'' in \emph{ISCA}, 2012.

\bibitem{cacti}
N.~Muralimanohar, R.~Balasubramonian, and N.~Jouppi, ``{Optimizing NUCA
  Organizations and Wiring Alternatives for Large Caches with CACTI 6.0},'' in
  \emph{MICRO}, 2007.

\bibitem{cuda-sdk}
{NVIDIA Corporation}, ``{CUDA C/C++ SDK Code Samples},''
  \url{http://developer.nvidia.com/cuda-cc-sdk-code-samples}, 2011.

\bibitem{fermi}
{NVIDIA Corporation}, ``{NVIDIA's Next Generation CUDA Compute Architecture:
  Fermi},''
  \url{http://www.nvidia.com/content/pdf/fermi_white_papers/nvidia_fermi_compute_architecture_whitepaper.pdf},
  2011.

\bibitem{kepler}
{NVIDIA Corporation}, ``{NVIDIA's Next Generation CUDA Compute Architecture:
  Kepler GK110},''
  \url{http://www.nvidia.com/content/PDF/kepler/NVIDIA-Kepler-GK110-Architecture-Whitepaper.pdf},
  2012.

\bibitem{maxwell}
{NVIDIA Corporation}, ``{NVIDIA GeForce GTX 750 Ti},''
  \url{http://international.download.nvidia.com/geforce-com/international/pdfs/GeForce-GTX-750-Ti-Whitepaper.pdf},
  2014.

\bibitem{mps}
{NVIDIA Corporation}, ``{Multi-Process Service},''
  \url{https://docs.nvidia.com/deploy/pdf/CUDA_Multi_Process_Service_Overview.pdf},
  2015.

\bibitem{pascal}
{NVIDIA Corporation}, ``{NVIDIA Tesla P100},''
  \url{https://images.nvidia.com/content/pdf/tesla/whitepaper/pascal-architecture-whitepaper.pdf},
  2016.

\bibitem{asplos-sree}
S.~Pai, M.~J. Thazhuthaveetil, and R.~Govindarajan, ``{Improving GPGPU
  Concurrency with Elastic Kernels},'' in \emph{ASPLOS}, 2013.

\bibitem{pichai-asplos14}
B.~Pichai, L.~Hsu, and A.~Bhattacharjee, ``{Architectural Support for Address
  Translation on GPUs: Designing Memory Management Units for CPU/GPUs with
  Unified Address Spaces},'' in \emph{ASPLOS}, 2014.

\bibitem{powers-hpca14}
J.~Power, M.~D. Hill, and D.~A. Wood, ``{Supporting x86-64 Address Translation
  for 100s of GPU Lanes},'' in \emph{HPCA}, 2014.

\bibitem{bip}
M.~K. Qureshi, A.~Jaleel, Y.~N. Patt, S.~C. Steely, and J.~Emer, ``{Adaptive
  Insertion Policies for High Performance Caching},'' in \emph{ISCA}, 2007.

\bibitem{ucp-yale}
M.~K. Qureshi and Y.~N. Patt, ``Utility-based cache partitioning: A
  low-overhead, high-performance, runtime mechanism to partition shared
  caches,'' in \emph{Proceedings of the 39th Annual IEEE/ACM International
  Symposium on Microarchitecture}.\hskip 1em plus 0.5em minus 0.4em\relax IEEE
  Computer Society, 2006.

\bibitem{fr-fcfs}
S.~Rixner, W.~J. Dally, U.~J. Kapasi, P.~Mattson, and J.~D. Owens, ``{Memory
  Access Scheduling},'' in \emph{ISCA}, 2000.

\bibitem{ccws}
T.~G. Rogers, M.~O'Connor, and T.~M. Aamodt, ``{Cache-Conscious Wavefront
  Scheduling},'' in \emph{MICRO}, 2012.

\bibitem{unitd}
B.~F. Romanescu, A.~R. Lebeck, D.~J. Sorin, and A.~Bracy, ``{UNified
  Instruction/Translation/Data (UNITD) Coherence: One Protocol to Rule them
  All},'' in \emph{HPCA}, 2010.

\bibitem{ptask}
C.~J. Rossbach, J.~Currey, M.~Silberstein, B.~Ray, and E.~Witchel, ``{PTask:
  Operating System Abstractions to Manage GPUs as Compute Devices},'' in
  \emph{SOSP}, 2011.

\bibitem{eaf-vivek}
V.~Seshadri, O.~Mutlu, M.~A. Kozuch, and T.~C. Mowry, ``{The Evicted-Address
  Filter: A Unified Mechanism to Address Both Cache Pollution and Thrashing},''
  in \emph{PACT}, 2012.

\bibitem{smith-hep}
B.~J. Smith, ``{A Pipelined, Shared Resource {MIMD} Computer},'' in
  \emph{ICPP}, 1978.

\bibitem{parboil}
J.~A. Stratton, C.~Rodrigues, I.~J. Sung, N.~Obeid, L.~W. Chang, N.~Anssari,
  G.~D. Liu, and W.~W. Hwu, ``{Parboil: A Revised Benchmark Suite for
  Scientific and Commercial Throughput Computing},'' {Univ.\ of Illinois at
  Urbana-Champaign}, Tech. Rep. { IMPACT-12-01}, March 2012.

\bibitem{gpuvm}
Y.~Suzuki, S.~Kato, H.~Yamada, and K.~Kono, ``{GPUvm: Why Not Virtualizing GPUs
  at the Hypervisor?}'' in \emph{USENIX ATC}, 2014.

\bibitem{isca-2014-preemptive}
I.~Tanasic, I.~Gelado, J.~Cabezas, A.~Ramirez, N.~Navarro, and M.~Valero,
  ``{Enabling Preemptive Multiprogramming on GPUs},'' in \emph{ISCA}, 2014.

\bibitem{cdc6600}
J.~E. Thornton, ``{Parallel Operation in the Control Data 6600},'' \emph{AFIPS
  FJCC}, 1964.

\bibitem{gVirt}
K.~Tian, Y.~Dong, and D.~Cowperthwaite, ``{A Full GPU Virtualization Solution
  with Mediated Pass-Through},'' in \emph{USENIX ATC}, 2014.

\bibitem{usui-squash}
H.~Usui, L.~Subramanian, K.~Chang, and O.~Mutlu, ``{SQUASH: Simple qos-aware
  high-performance memory scheduler for heterogeneous systems with hardware
  accelerators},'' \emph{arXiv CoRR}, 2015.

\bibitem{usui-dash}
H.~Usui, L.~Subramanian, K.~Chang, and O.~Mutlu, ``{DASH: Deadline-Aware
  High-Performance Memory Scheduler for Heterogeneous Systems with Hardware
  Accelerators},'' \emph{ACM TACO}, vol.~12, no.~4, Jan. 2016.

\bibitem{abhishek-ispass16}
J.~Vesely, A.~Basu, M.~Oskin, G.~H. Loh, and A.~Bhattacharjee, ``{Observations
  and Opportunities in Architecting Shared Virtual Memory for Heterogeneous
  Systems},'' in \emph{ISPASS}, 2016.

\bibitem{vijay-hpca17}
T.~Vijayaraghavany, Y.~Eckert, G.~H. Loh, M.~J. Schulte, M.~Ignatowski, B.~M.
  Beckmann, W.~C. Brantley, J.~L. Greathouse, W.~Huang, A.~Karunanithi,
  O.~Kayiran, M.~Meswani, I.~Paul, M.~Poremba, S.~Raasch, S.~K. Reinhardt,
  G.~Sadowski, and V.~Sridharan, ``{Design and Analysis of an APU for Exascale
  Computing},'' in \emph{HPCA}, 2017.

\bibitem{vmCUDA}
L.~Vu, H.~Sivaraman, and R.~Bidarkar, ``{GPU Virtualization for High
  Performance General Purpose Computing on the ESX Hypervisor},'' in
  \emph{HPC}, 2014.

\bibitem{wang-hpca16}
Z.~Wang, J.~Yang, R.~Melhem, B.~R. Childers, Y.~Zhang, and M.~Guo,
  ``{Simultaneous Multikernel GPU: Multi-tasking Throughput Processors via
  Fine-Grained Sharing},'' in \emph{HPCA}, 2016.

\bibitem{demystify}
H.~Wong, M.-M. Papadopoulou, M.~Sadooghi-Alvandi, and A.~Moshovos,
  ``{Demystifying GPU Microarchitecture Through Microbenchmarking},'' in
  \emph{ISPASS}, 2010.

\bibitem{xie-iccad13}
X.~Xie, Y.~Liang, G.~Sun, and D.~Chen, ``{An Efficient Compiler Framework for
  Cache Bypassing on {GPUs}},'' in \emph{ICCAD}, 2013.

\bibitem{xie-hpca15}
X.~Xie, Y.~Liang, Y.~Wang, G.~Sun, and T.~Wang, ``{Coordinated Static and
  Dynamic Cache Bypassing for {GPUs}},'' in \emph{HPCA}, 2015.

\bibitem{warp-slicer}
Q.~Xu, H.~Jeon, K.~Kim, W.~W. Ro, and M.~Annavaram, ``{Warped-Slicer: Efficient
  Intra-SM Slicing through Dynamic Resource Partitioning for GPU
  Multiprogramming},'' in \emph{ISCA}, 2016.

\bibitem{pagoda-ppopp17}
T.~T. Yeh, A.~Sabne, P.~Sakdhnagool, R.~Eigenmann, and T.~G. Rogers, ``{Pagoda:
  Fine-Grained GPU Resource Virtualization for Narrow Tasks},'' in
  \emph{PPoPP}, 2017.

\bibitem{complexity}
G.~Yuan, A.~Bakhoda, and T.~Aamodt, ``{Complexity Effective Memory Access
  Scheduling for Many-Core Accelerator Architectures},'' in \emph{MICRO}, 2009.

\bibitem{tianhao-hpca16}
T.~Zheng, D.~Nellans, A.~Zulfiqar, M.~Stephenson, and S.~W. Keckler, ``{Towards
  High Performance Paged Memory for GPUs},'' in \emph{HPCA}, 2016.

\bibitem{frfcfs-patent}
W.~K. Zuravleff and T.~Robinson, ``{Controller for a Synchronous DRAM That
  Maximizes Throughput by Allowing Memory Requests and Commands to Be Issued
  Out of Order},'' in \emph{US Patent Number 5,630,096}, 1997.

\end{thebibliography}

\end{document}